\documentclass[aps,12 pt, showpacs, eqsecnum]{revtex4}
\usepackage{amsmath}
\usepackage{amsfonts}
\usepackage{amssymb}
\usepackage{color}

\def\bc{\begin{center}}
\def\nno{\nonumber}
\def\ec{\end{center}}
\def\be{\begin{eqnarray}}
\def\ee{\end{eqnarray}}

\newcommand{\omits}[1]{}

\definecolor{dyellow}{rgb}{1.,0.8,.0}
\definecolor{myblue}{rgb}{.1,.1,.7}
\definecolor{dcyan}{rgb}{.0,.6,.6}
\definecolor{dmagenta}{rgb}{0.6,0.0,0.6}
\definecolor{brown}{rgb}{0.6,0.2,0.}
\definecolor{darkblue}{rgb}{.0,.0,0.5}
\definecolor{darkred}{rgb}{0.75,0.0,0.0}
\definecolor{orange}{rgb}{1.,.6,.0}
\definecolor{dorange}{rgb}{0.8,.4,.0}
\definecolor{darkgreen}{rgb}{0.0,0.6,0.0}
\definecolor{purple}{rgb}{.4,.0,.4}

\def\red{\color{red}}
\def\blue{\color{blue}}

\def\La{\Lambda}
\def\Si{\Sigma}

\def\a{\alpha}
\def\b{\beta}

\def\dl{\delta}

\def\si{\sigma}

\def\d#1#2{\frac{\displaystyle #1}{\displaystyle #2}}

\newcommand{\dS}{$d{S}$}
\newcommand{\AdS}{${A}dS$}
\newcommand{\BdS}{${B}d{S}$}
\newcommand{\BAdS}{${BA}d{S}$}

\newcommand{\HlsM}{${H}_{R+} \subset M^{1,4}$}
\newcommand{\HsM}{${H}_{+} \subset M^{1,4}$}
\newcommand{\HrsM}{${H}_R \subset M^{1,4}$}

\newcommand{\GR}{general relativity}
\newcommand{\SR}{special relativity}

\newcommand{\PoE}{principle of equivalence}

\newcommand{\PoR}{principle of relativity}
\newcommand{\PoI}
{principle of inertia}
\newcommand{\CP}{cosmological principle}
\newcommand{\CFT}{$CFT$}
\newcommand{\Mink}{${M}ink$}

\newcommand{\PoL}
{principle of localization}

\newcommand{\FLT}{$FLT$}

\newcommand{\E}{$E$}

\newcommand{\CG}{$SO(2,4)$}
\newcommand{\cg}{$\mathfrak{so}(2,4)$}
\newcommand{\NsM}{${\cal N}\subset M^{2,4}$}

\newcommand{\vect}[1]{\ensuremath{\boldsymbol{#1}}}

\newcommand\btd{\raise 2pt
\hbox{$\hat\bigtriangledown$}\hskip 1.5pt}
\newcommand\bt{\raise 2pt
\hbox{$\bigtriangledown$}\hskip 1.5pt}

\def\PRD{{\it Phys. Rev.}~{\bf D}}

\def\PLA{{\it Phys. Lett.}~{\bf A}}

\def\CTP{{\it Commun. Theor. Phys. }}

\begin{document}
\title{\Large Special Relativity and Theory of Gravity \\[-2mm] via \\[-2mm]
 Maximum Symmetry and Localization\\[1mm]

\large --- In Honor of the 80th Birthday of Professor Qikeng Lu
\footnote{Invited talk at the
 2006 International Conference on Several Complex Variables, June 5 - 9, 2006; Beijing,
 China.}}
\author{Han-Ying Guo\thanks{Email: \tt hyguo@itp.ac.cn}}

\affiliation{%
Institute of Theoretical Physics, Chinese Academy of Sciences,
Beijing 100080, China.}


\begin{abstract}
Like Euclid, Riemann and Lobachevsky geometries are on an almost
equal footing, based on the principle of relativity of maximum
symmetry proposed by Professor Qikeng Lu and the postulate on
invariant universal constants  $c$ and $R$, the de Sitter/anti-de
Sitter (\dS/\AdS)
 special relativity on  \dS/\AdS-space with radius $R$
can be set up on an almost equal footing with Einstein's special
relativity on Minkowski-space as the case of $R \to \infty$.

Thus the \dS-space is coin-like: A law of inertia in Beltrami atlas
with Beltrami time simultaneity for the \PoR\ on one side. The
proper-time simultaneity and a Robertson-Walker-like \dS-space with
entropy and an accelerated expanding $S^3$ fitting the cosmological
principle on another.

If our universe is asymptotic to the Robertson-Walker-like \dS-space
of $R\backsimeq (3/\Lambda)^{1/2}$, it should be slightly closed in
$O(\Lambda)$
 with entropy bound $S\simeq3\pi {c^3k_B/\Lambda
G\hbar}$. Contrarily, via its asymptotic behavior, it can fix on
Beltrami inertial frames without  `an argument in a circle' and acts
as the origin of inertia.

 There is a triality of conformal extensions
of three kinds of \SR\ and their null physics on the projective
boundary of a 5-d \AdS-space, a null cone modulo projective
equivalence $[{\cal N}]\cong\partial_P({A}d{S}^5)$. Thus there
should be a \dS-space on the boundary of $S^5 \times AdS^5$ as a
vacuum of supergravity.

In the light of Einstein's `Galilean regions', gravity should be
based on the localized \PoR\ of full maximum symmetry with a
gauge-like dynamics. Thus, this may lead to theory of gravity of
corresponding {local symmetry}.
 A simple model of \dS-gravity characterized by a dimensionless constant $g\simeq (\Lambda G\hbar
/3c^{3})^{1/2}\sim 10^{-61}$
 shows the
features on umbilical manifolds of local
 \dS-invariance. Some gravitational
 effects out of \GR\,  may play a role as  dark matter.

The dark universe and its asymptotic behavior may already indicate
that the \dS\ \SR\ and \dS-gravity
  be the foundation  of large scale
physics.
\end{abstract}


\maketitle

\tableofcontents

\section{Introduction}

As a famous mathematician, Professor Qikeng Lu's contributions to
physics concern various fields, such as dispersion relations,
special and general relativity, theory of gravity, gauge theory,
integrable systems, conformal field theory, so on and so forth. As
one of his great contributions, he has suggested that the
principle of relativity should be generalized to constant
curvature spacetimes with radius $R$, i.e. de Sitter (\dS) and
anti-de Sitter (\AdS) spacetimes \cite{Lu} and began to research
the special relativity on these maximally symmetric spacetimes in 1970s
\cite{Lu, LZG}.
 Actually, based on the \PoR\  of
maximum symmetry 
 and the
 postulate on invariant universal constants of $c$ and $R$ \omits{\blue\cite{BdS, BdS2}}, 
 the \dS/\AdS-invariant special
 relativity, the \dS/\AdS\ \SR\ for short, can be set
 up \cite{Lu, LZG, lu05,
BdS, BdS2, BdS3, IWR, TdS, NH, yan, duality, OoI, PoI, dual07} on
an almost equal footing with Einstein's \SR\ on
Minkowski(\Mink)-spacetime  as the case of $R \to \infty$.

As is well known, Einstein's theory of relativity including
special relativity, general relativity and cosmology provides some
of most important breakthroughs in the last century. Together with
quantum theory, they constitute the foundation of modern physics
although how to quantize gravity formulated in  general relativity
is still open. Recent observations \cite{Riess98, WMAP} show,
however, that our universe is almost completely dark and
accelerated expanding. It is not asymptotic to a \Mink-spacetime,
but possibly a \dS-spacetime with a tiny positive cosmological
constant $\Lambda>0$. With plenty of \dS-puzzles, these greatly
challenge Einstein's theory of relativity as a foundation of
physics in large scale.

The cosmological constant is regarded as some quantum `vacuum'
energy in ordinary approach. This leads to a huge difference of
$10^{-122}$ as real $\Lambda>0$ is extremely tiny. According to
Professor Lu's proposal \cite{Lu} and the \dS\ \SR\ \cite{Lu, LZG,
lu05, BdS, BdS2, BdS3, IWR, TdS, NH, yan, duality, OoI, PoI,
dual07}, however, it should be one of the fundamental constants in
the Nature like the speed of light $c$, Newton's gravitational
constant $G$ and Planck constant $\hbar$. Thus, the huge
difference puzzle should transfer to another issue.

Why there should be three kinds of \SR\ with maximum symmetry?

When Poincar\'e first introduced the \PoR\ as one of the most
important principles in the Nature\cite{1904}, he inherited the
 assumption  from Newton that  space and time be Euclidean.
In his first paper on  \SR\ \cite{1905}, Einstein also took this
assumption and required that a rest rigid ruler be Euclidean. But,
there is nowhere exact flat in either our universe or its
asymptotical region except in the sense of Einstein's `Galilean regions'
\cite{1923} where gravity and the dark energy  can be completely
ignored. Actually, just like weakening Euclid's fifth axiom leads to
non-Euclidean geometry, giving up the Euclidean assumption should first lead
to  two other kinds of \dS/\AdS\ special relativity \omits{with
$SO(1,4)/SO(2,3)$-invariance on \dS/\AdS-spacetime, respectively
\cite{Lu, LZG, lu05, BdS, BdS2, BdS3, IWR, TdS, NH, yan, duality,
OoI, PoI}, }on an almost equal footing
 with
Einstein's special relativity. 

In geometry, Euclid, Riemann and Lobachevsky geometries as three
classes of constant curvature ones of maximum symmetries, there
are Descartes coordinate systems for Euclid geometry or Beltrami
coordinate systems \cite{beltrami} for non-Euclidean ones (see
also \cite{ klein1871, R, hpg97})\omits{\footnote{For the sake of
simplicity and more understandable, we employ the coordinate
description. Needless to say, some coordinate-free description can
also be set up formally.}} and all geodesics in these systems are  simultaneously
straight lines of linear form, respectively. These systems
in four dimensions, for example, with points, straight lines and
metric symmetrically transformed under the linear transformations of
$ISO(4)$ for Euclid geometry or the fractional linear ones with a
common denominator (\FLT s) of $SO(5)$ for Riemann geometry and of
$SO(1,4)$ for Lobachevsky geometry, respectively. Beltrami
\cite{beltrami} introduced such coordinate %
in order to show the consistency of Lobachevsky plane. It is
completed by Klein \cite{klein1871}. %

Changing signature by a Weyl unitary trick or
 an inverse Wick rotation, these spaces with corresponding coordinate systems
  become
 $ISO(1,3)$/$SO(1,4)$/$SO(2,3)$-invariant \Mink/\dS/\AdS-spacetime with
  \Mink-systems and  Beltrami  systems,
respectively \cite{IWR}. At the same time,  points, geodesics being
straight lines and metrics turn out to be  events, geodesics being
straight world-liners and \Mink\ or Beltrami metric of
physical signature 
 in
relevant coordinate systems symmetrically transformed under
corresponding transformations of group
$ISO(1,3)$/$SO(1,4)$/$SO(2,3)$, respectively. In analogy with those
on  \Mink-spacetime, the motions along straight world-lines and the
  Beltrami systems on \dS/\AdS-spacetime should be of inertia.
Thus, there should be a law of inertia and a \PoR\ on
\dS/\AdS-spacetime, respectively. All these properties are also
true globally on the \dS/\AdS-spacetime with  Beltrami coordinate
atlas.

 As was claimed by Klein: `Geometry of
space is associated with mathematical group'\cite{klein1872}, the
idea of invariance of geometry under transformation group may
imply that on some spacetimes of maximum symmetries there should be a \PoR,  which requires the invariance of physical laws
without
 gravity under transformations
 among inertial systems.
 This is just the key point of  Lu's
proposal to generalize the invariance of maximally symmetry for
physical laws without gravity
 to all maximally symmetric spacetimes \cite{Lu}. \omits{In analogy with that Einstein's special
relativity is set up on the \Mink-spacetime as a counterpart of
Euclid, the counterparts of two kinds of non-Euclidian geometries
 are just  the \dS/\AdS\ \SR\
on the \dS/\AdS-spacetime, respectively. }Further, all other kinds
of \PoR\ on  corresponding space-time, such as Galilei \PoR\ on
Newton's space and time, Newton-Hooke/anti-Newton-Hooke \PoR\
\cite{NH} on Newton-Hooke/anti-Newton-Hooke space-time and even
Poincar\'e \PoR\ on \Mink-spacetime, can be regarded as certain
contraction of the \dS/\AdS-invariant \PoR\ on \dS/\AdS-spacetime in
different limiting case, respectively. Thus, the significance of
Lu's
proposal 
for   relativistical physics is more or less like Klein's Erlangen
program for  geometry.

In the \dS\ \SR, there are some very important issues.

For free particles and
light signals, in addition to the law of inertia there is  
a set of conserved observables  with a generalized Einstein's
formula on
mass-energy-momentum-boost-angular momentum. %
The famous Einstein's formula on mass-energy-momentum is the case of $R\to
\infty$.

 There are two kinds of simultaneity. For  the \PoR\ with inertial
 observers and
  inertial law,
 there is Beltrami time
simultaneity.
 The proper-time simultaneity is for these observers' comoving-like
observations. If the proper-time is taken as a temporal
coordinate, inertial observers become  comoving-like ones and the
Beltrami metric transfers to its Robertson-Walker-like \dS\
counterpart with  an accelerated expanding closed 3-cosmos $S^3$,
which fits the cosmological principle with \dS-symmetry.
 Thus, the \dS-spacetime with both the \PoR\ and the \CP\ is just
 like  a coin with two sides. Actually,
 the maximum symmetry ensures that these
principles do  make sense in  different  sides of the coin. Thus,
the Robertson-Walker-like \dS-cosmos
 acts as the
origin of the inertial law in  Beltrami inertial frames.  And  the
\PoR\ with the inertial law  on  Beltrami metric provides a
benchmark for physics on \dS-spacetime. Further qualitatively, due
to the generalized Einstein formula and the \dS-symmetry, all free
moves of test objects such as celestial objects including the cosmic
microwave background radiation (CMB)  as a whole  should have both
conserved energy-momentum and angular momentum.

For the \dS-horizon puzzle, i.e. why the \dS-spacetime of constant
curvature is like a black hole, there is another explanation. The
\dS-horizon in  Beltrami systems  is actually at $T=0$ without
entropy, while at the horizon in other systems, such as the static
\dS-universe and the Robertson-Walker-like \dS-spacetime, Hawking
temperature and area entropy appear as non-inertial effects rather
than that of gravity \cite{TdS}. Thus,  \dS-spacetime is completely
different from a black hole.

 {\blue Since} our universe
is asymptotic to {\blue a \dS-spacetime, which should be} such a
Robertson-Walker-like \dS-space {\blue with }\omits{ entropy and}
$R\simeq (3/\Lambda)^{1/2}$, it should be an evolutional slightly
closed \fbox{3-dimensional} cosmos curved in the order of $\Lambda$,
$O(\Lambda)$, with an upper entropy bound. The closeness with very
tiny curvature of our universe is a simple but important prediction.
It is more or less indicated by the data from WMAP recently
\cite{WMAP}. \omits{In addition, since that Beltrami systems  are of
inertia does not depend on the concrete values of $c$ and $R$, }On
the other hand, the evolution of our universe can fix on a kind of
Beltrami inertial frames  via their Robertson-Walker-like \dS\
counterpart as the fate of our universe \cite{PoI}. Thus, for the
principle of inertia on \dS-space  there is no Einstein's `argument
in a circle' \cite{1923}. Further, if all other kinds of \PoI\ are
regarded as contractions under different limits of the \PoI\ on
\dS-space, this is also true for all kinds of \PoI.

A scaling of $R$  leads to  conformal extensions
of the \dS/\AdS\ \SR\  on \dS/\AdS-spacetime. 
Together with  conformal extension of Einstein's \SR\ on
\Mink-spacetime, in fact, all these conformal extensions
 are on a
null cone modulo projective equivalence isomorphic to the
projective boundary of a 5-dimensional  \AdS-spacetime, $[{\cal
N}]\cong
\partial_P({A}d{S}^5)$. Thus,
 there is a triality of  conformal
extensions of three kinds of \SR\ and   null physics on
\Mink/\dS/\AdS-spacetimes. And certain Weyl mappings relate any two
of them \cite{c3}. {\blue Further,  there should be a \dS-spacetime
on the boundary of $S^5 \times AdS^5$ as a vacuum of supergravity.}

According to \GR, there is no \SR\ on \dS/\AdS-spacetime. Different from 
 \GR, in  view of  the \dS/\AdS\ \SR, there is no gravity on
\dS/\AdS-spacetime. We should explain how to describe gravity in
the universe.

In the light of Einstein's `Galilean regions' where his \SR\, should
hold locally since the regions are essentially `finite' \cite{1923},
it is the core of Einstein's idea on  spacetime with gravity that it
should be curved with localized \SR\ of local full Poincar\'e
symmetry \cite{PoI}. In Einstein's general relativity, however,
there are local Lorentz frames of only local $SO(1,3)$ invariance
rather than full local Poincar\'e invariance with local
translations. Thus, in Einstein's \GR, the benchmark of physics for
defining physical quantities and introducing laws of physics with
gravity is not completely in consistency with that in Einstein's
\SR\ \cite{PoI}. In addition, there is a `Gordian knot' in dynamics
(see, e.g., \cite{MTW}). These may cause some puzzles.

Taking into account the localization of special relativity, theory
of gravity should be based on  a generalized equivalence principle
with full localized maximum symmetry of \SR\  called the principle
of localization. In consistency with this principle, it can be
further  expected that gravity  be governed by a
gauge-like dynamics with  same local maximum symmetry. 
Thus, the localization of three kinds of special relativity leads to
three kinds of theories of gravity\omits{with relevant local special
relativity and gauge-like dynamics} with  full local maximum
symmetry. The Nature {\blue should} prefer one of them.

How to realize mathematically the localization of three kinds of
special relativity?

It is needed to localize \Mink/\dS/\AdS-spacetime as maximally
symmetric spacetime $\cal S$ with  maximum symmetry group $\mathfrak
G$ and to {\blue patch} them together as a kind of differential
manifolds $({\cal M}, g, \Gamma)$ with  metric $g$ and  metric
compatible connection $\Gamma$ valued in  Lie algebra $\mathfrak g$
of $\mathfrak G$. That is, in terminology of
 fibre bundle and connection theory \cite{KN}, it is needed to set up a principal bundle
  $P({\cal M}, {\mathfrak G})$ over such an
$\cal M$ with  $\mathfrak G$ as a structure group and  an
associated bundle $E({\cal M}, {\cal S}, {\mathfrak G}, P)$ with
$\cal S$ as a typical fibre. In addition, there should be some
associated bundles with certain irreducible representations of
$\mathfrak G$  as fibre fitted by the matter fields as sources,
and so on. Since  transformations on the  Beltrami model of
 \dS-spacetime are of \FLT s,  these requirements
may lead to some connection valued in $\mathfrak g$ realized
non-linearly. In fact, this is one of motivations for Lu to study
the non-linear connection theory \cite{NLC}.

There are still some physical issues to be precisely set up for
such  geometric description of  spacetimes with gravity based on
the principle of localization, such as the relation between the
metric with local maximum symmetry $\mathfrak G$ and the
connection valued in $\mathfrak g$ and so on. However, there is a
simple model of \dS-gravity with a
 gauge-like action   characterized by a dimensionless
constant $g \simeq (\Lambda G\hbar
  /3c^{3})^{1/2}\sim 10^{-61}$ \cite{dSG, uml, T77, QG}
   on a
kind of umbilical Riemann-Cartan manifolds with local \dS-invariance
\cite{uml}. It has partially shown these features and may present an
explanation of \fbox{the} dark matter in terms of the gravitational
 effects out of \GR, at least partially. Therefore, it may  provide  an alternative framework for  data analysis in
precise cosmology.

It should be notes that  $g^2$ is in the same order of the huge
difference of $\Lambda$ as so-called  quantum `vacuum energy'. Then
there are further questions: What is the origin of the dimensionless
constant $g$? Is it with other dimensionless constants calculable?

This paper is arranged as follows. In section \ref{RvS},  we explain why there should be two other  kinds of
 special  relativity with \dS/\AdS-invariance by means of  the
 Beltrami  model  of  Riemann sphere and its physical
counterpart on \dS-spacetime. Some historical remarks are also made.\omits{ on
the Beltrami model of \dS-spacetime.} In section \ref{dSSR}, we
briefly introduce the properties and  cosmological significance of
the \dS\ \SR.  We explain how the evolution of our universe can fix
on the inertial systems without Einstein's `argument in a circle'.
In section \ref{C3}, we show that there is a triality of conformal
extensions of three kinds of \SR\ and  null physics on 4-dimensional
\Mink/\dS/\AdS-spacetimes on the projective boundary of a
5-dimensional \AdS-spacetime. In section \ref{gravity}, in the light
of Einstein's `Galilean regions' in spacetime with gravity, we explain why gravity should be
based on the principle of
localization, i.e. on the localized  special relativity 
with full maximum symmetry. 
We also  briefly introduce the simple model of \dS-gravity with an
action of gauge-like on  umbilical manifolds. Finally, we {end}
with some {concluding} remarks.

\section{Three Kinds of Special Relativity with Maximum Symmetry}\label{RvS}%

\subsection{Beltrami model of  Riemann sphere}


Let us
 focus on the {Beltrami}  model of Riemann sphere ${S}_R^4$ \cite{IWR, dual07}, 
 since its physical counterpart is just  the {Beltrami}  model of
 \dS-spacetime, denoted \BdS-space.
  Similarly, we may consider the Beltrami model of Lobachevsky hypernoloid $L^4$ and
 the one of \AdS-spacetime as its physical
  counterpart. In the original
  Beltrami model \cite{beltrami, klein1871}  for Lobachevsky  geometry, it is
 of one coordinate chart (see, e.g.
 \cite{R, hpg97}).  For the case of Riemann sphere\fbox{$S^4$}, however, one chart is not
 enough. But, 
 all fundamental properties 
 can be
 generalized to a  Beltrami
 coordinate atlas covering the sphere chart by chart.

A Riemann sphere ${S}_R^4$ with radius $R$ can be embedded in a
5-dimensional  Euclid space \E$^5$
\be\label{4s}%
{S}_R^4:~&&\delta_{AB}\xi^A \xi^B=R^2>0, \quad A, B=0, \cdots, 4,\\\label{5ds}%
&& ds_E^2=\delta_{AB}d\xi^A d\xi^B=d\xi {I}d\xi^t,
\ee%
where $I=(\delta_{AB})=diag(1,\cdots, 1),~ \xi=(\xi^0,\cdots,
\xi^4)$. They are invariant under (linear) rotations:
\be\label{so5}%
 \xi~ \rightarrow ~\xi'= \xi S, \quad S {I} S^t={I}, ~~\forall S \in {SO(5)}.
\ee%
\omits{They are invariant under (linear) transformations 
of 5-dimensional  rotation group $SO(5)$.}

In the surface theory, an $S^2 \subset E^3$ is an umbilical one
(see, e.g., \cite{Spivak}). This is also the case for the Riemann
sphere $S_R^4 \subset E^5$.

The Beltrami model provides an intrinsic geometry of ${S}_R^4$ on
the
Beltarmi-space 
${B}_R \cong { S}_R^4$ 
with Beltrami coordinates atlas. All 
properties of ${S}_R^4$ are well-defined in the  atlas.  For an
orientable intrinsic geometry of ${B}_R$, it is needed an atlas
with eight charts: $U_{\pm a}:= \{ \xi\in{ S}_R^4 : \xi^a\gtrless
0\}, a=1, \cdots, 4$. In the chart $U_{+4}$, for instance, the
Beltrami coordinates are
\be \label{u4}%
x^i|_{U_{+4}} =R \frac{\xi^i}{\xi^4},\qquad i=0,\cdots, 3;\quad
\xi^4|_{U_{+4}}> 0.
\ee%
 In another chart $U_{+3}$, say,
\begin{equation}  %
y^{j'}|_{U_{+3}}=R\frac{\xi^{j'}}{\xi^3},\quad
j'=0,\cdots,\hat{3}\cdots,4; \quad \xi^{3}|_{U_{+3}}> 0,
\end{equation}
where $\hat{3}$ means omission of $3$. It is important that the
transition function $T_{+4,+3}$ on the intersection  $U_{+4}\cap
U_{+3}$ is of \FLT : $T_{+4,+3} =\xi^3/\xi^4=x^3/R=R/y^4$ so that
$x^i=T_{+4,+3}y^{i'} $  for $i=i'=0,1,2$ and $x^3=R^2/y^4$ are of
\FLT s. %

 In the
chart $U_{+4}$, say, \omits{ i.e. the inhomogeneous coordinates in
projective geometry \cite{Rosenfeld, beltrami} (hereafter, we call
them after Beltrami
for short),%
\be\label{Bcrd}%
x^i:=R\frac{\xi^i}{\xi^4},
\quad \xi^4\neq 0  ,\quad
 i=0,\cdots, 3.%
\ee%
T}Riemann sphere (\ref{4s}) and   metric (\ref{5ds}) 
\omits{%
ds_E^2=\delta_{AB}d\xi^A d\xi^B %
\ee}%
 restricted
 on ${B}_R$ becomes   domain condition and  Beltrami metric, 
 respectively:
 \be\label{sigma}%
 {B}_R:&& \sigma_E(x):=\sigma_E(x,x)=1+R^{-2}\delta_{ij}x^ix^j>0,\\\label{4Bds}%
&&ds_E^2=\{\delta_{ij}\sigma_E^{-1}(x)-R^{-2}\sigma_E^{-2}(x)\delta_{il}x^l\delta_{jk}x^k\}
 dx^idx^j, 
 \ee
which are invariant under the \FLT s among
Beltrami coordinates $x^i$ 
in a transitive form  sending a point $A(a^i)$ to the origin
$O(o^i=0)$,%
 \be\nno
x^i\rightarrow
\tilde{x}^i&=&\pm\sigma_E(a)^{1/2}\sigma_E(a,x)^{-1}(x^j-a^j)N_j^i,\\\label{FLT}
N_j^i&=&O_j^i-{ R^{-2}}%
\delta_{jk}a^k a^l (\sigma_E(a)+\sigma_E(a)^{1/2})^{-1}O_l^i,\\\nno
O&:=&(O_j^i)_{i,j=0,\cdots,3}\in SO(4).%
 \ee 

There is an invariant for two points $A(a^i)$ and $X(x^i)$ on
${B}_R$, which corresponds to the cross ratio
among the  points together with the origin and the infinity in projective geometry approach: 
\be\label{AB} %
{\Delta}_{E,R}^2(a, x) = R^2
[\sigma_E(a)\sigma_E(x)-\sigma_E^2(a,x)].
\ee %
For two  adjacent points $X(x^i)$ and $X'(x^i+dx^i)$, this
invariant is just the Beltrami metric (\ref{4Bds}).

The proper length between $A(a^i)$ and $B(b^i)$ is an integral of
$ds_E$ over the geodesic segment $\overline{AB}$:
\be \label{ABL}%
L_E(a,b)&=& R \arcsin (|\Delta_E(a,b)|/R).
\ee

As was mentioned, there is an important property in the model: All
geodesics of the Beltrami metric are  straight lines  linearly. This
property of the Beltrami coordinates is different from other
coordinates for the Riemann sphere and also different from other
non-maximally symmetric spaces in Riemannian differential geometry
in general.

In fact, the geodesics of the Beltrami metric are equivalent to %
\be\label{geod}%
\frac{d q^i}{ds_E}=0,\quad{q^i}:=\sigma_E^{-1}(x)\frac{dx^i}{ds_E}.%
\ee%
 Therefore, 
\be\label{q}%
{q^i}
={ consts}.%
\ee%
Further, it is easy to see that the following rations are constants %
\be\label{um}%
\frac{q^\alpha}{q^0}=\frac{dx^\alpha}{dx^0}=consts, \quad \alpha=1,2,3.%
\ee%
The eqn. (\ref{geod})  can be integrated further to get the linear
result:
\be\label{sl}%
x^i(s)=\alpha^i x^0(s)+\beta^i;\quad \alpha^i,\beta^i=consts.%
\ee%
Under the \FLT s (\ref{FLT}) among Beltrami systems, all these properties 
are transformed among themselves. They are also well established
globally chart by chart. 

 From the  viewpoint of  projective geometry,
 Beltrami coordinates are similar to  inhomogeneous projective ones
 and  antipodal identification should not be taken in order to preserve
orientation.


\omits{What is the physics on \BdS$_R$\ in spacetime? Needless to
say, this is a extreme important question.}

 \subsection{ Beltrami model of \dS-spacetime}
  Via a Weyl unitary trick or an inverse Wick rotation of $E^5$, which turns $\xi^0$ to be
time-like, the
  Riemann sphere ${S}_R^4\subset E^5$ and its Beltrami model  ${B}_R$
  becomes the \dS-hyperboloid \HlsM\ and its \BdS-model, a \dS-spacetime with
  the Beltrami atlas, respectively \cite{BdS, BdS2, BdS3, IWR}. In fact, a Weyl trick or an inverse Wick rotation changes all
$(\delta_{AB})$,
$(\delta_{ij})$ in all  metrics %
to $ (\eta_{AB}):=diag(1,-1,\cdots,-1)$, $(\eta_{ij}):=diag(1,-1,-1,-1)$ %
 and the sign of $R^2$ in all formulas. %
Then both $\xi^0$ and $x^0$ become time-like. In order to introduce
the time coordinate, a universal constant $c$ of speed dimension is
needed, say $x^0=ct$. Thus, there are two universal constants $c$
and $R$.

  Let us
  briefly review the \dS-hyperboloid, its \BdS-model and some  physics on them.
 \subsubsection{\dS-hyperboloid \HlsM\ and uniform `great circular' motion}

  The  \dS-hyperboloid can be embedded in a $4+1$-dimensional \Mink-spacetime \HlsM\ or simply \HsM:
\be\label{H+}%
 { H}_{R+}:~& &\eta_{AB}\xi ^{A}\xi ^{B}\omits{-\sum
_{\alpha=1}^{3}\xi ^{\alpha}\xi ^{\alpha}-\theta\xi ^{4}\xi ^{4}}=-
R^2<0, ~~\small
A,B=0,\cdots,4,
\\\label{dsH} %
&&ds_+^2=\eta_{A B}d\xi^A d\xi^B=d\xi J d\xi^t, \\\label{bdryH}
\partial_P H_{R+}: &&\eta_{AB}\xi^A\xi^B=0,%
\ee
where ${J}=(\eta_{A B})=
diag (1,-1,-1,-1,-1 )$, 
$\partial_P$ the projective boundary.
They  are invariant under (linear) transformations of 
\dS-group $SO(1,4)$:
\be\label{SD14}%
\xi~ \rightarrow ~\xi'=\xi S, \quad S {J} S^t={J},\quad \forall S
\in {SO(1,4)}.
\ee%
%

It is clear that the \dS-hyperboloid 
\HsM\  (\ref{H+}) is also an umbilical hypersurface of constant
curvature in the following sense: At any given point $\forall P \in$
\HsM, the first and second fundamental forms are proportional to
each other with a coefficient $R$. In addition, there are a tangent
\Mink-space $T_P(H_+)$  at $P$ and a radius vector ${\bf r}_P$
opposite to the normal vector with respect to the tangent space,
i.e. ${\bf r}_P=-{ (N=Rn)}_P$, where $n_P$ is a unit base of the
normal space $N^1_P$, and $T_P(H_+)\times {N}^1_P \cong M^{1,4}$.
This structure will be useful for the localization of the \HsM.

Corresponding to  great circles as  geodesics on the Riemann
sphere ${S}_R^4$, there  should be a kind of uniform `great
circular' motions for a particle with mass $m_R$
 along  geodesics on the \dS-hyperboloid \HsM\ defined by a conserved 5-dimensional  angular momentum
 ${L}^{AB}$: 
\be\label{angular5}%
  \frac{d{L}^{AB}}{ds_+}=0, \qquad {L}^{AB}:=m_R(\xi^A\frac{d\xi^B}{ds_{+}}
  -\xi^B\frac{d\xi^A}{ds_{+}}).%
 \ee%
And for the particle, there is  an
Einstein-like formula%
\begin{eqnarray}\label{5eml}%
{ -\frac{1}{2R^2}{L}^{AB}{L}_{AB}=m_R^2},\quad
{L}_{AB}=\eta_{AC}\eta_{BD}{L}^{CD}.
\end{eqnarray}

For a massless particle or a light signal with $m_R=0$, similar
uniform `great circular' motion can also be defined as long as the
proper-time in $ds_{+}$ is replaced by an affine parameter
$\lambda_+$ and there is no $m_R$ in the counterpart of ${
L}^{AB}$ in (\ref{angular5}), respectively. Namely,
\be\label{L5}%
 \frac{d{
L}^{AB}}{d\lambda_+}=0,\quad {
L}^{AB}:=\xi^A K^B
-\xi^B K^A, \quad K^A:=\frac{d\xi^A}{d\lambda_+}.%
\ee%
There is also an Einstein-like formula for the massless case.\omits{%
\begin{eqnarray}\label{eml5}%
-\frac{1}{2R^2}{ L}^{AB}{ L}_{AB}=0,\quad%
{ L}_{AB}=\eta_{AC}\eta_{ BD}{L}^{CD}.
\end{eqnarray}}

  In order to make sense for these motions,
 simultaneity should be defined.

 There are two time-like scales
 on the \dS-hyperboloid, the coordinate-time $\xi^0$ and the proper-time $s_+$.
For a pair of  events $(P(\xi_P), Q(\xi_Q))$, they are
simultaneous in the coordinate-time if and only if%
\be\label{simtH}%
 \xi_P^0=\xi_Q^0.%
 \ee%
A simultaneous 3-hypersurface of $\xi^0=const$ is an expanding $S^3$：%
\be\label{s3}%
\delta_{ab}\xi^a\xi^b&=&R^2+(\xi^0)^2,~~ a, b=1,\cdots, 4;\\\nonumber
dl^2|_{\xi^0=const}&=&\delta_{ab}d\xi^a d\xi^b. %
\ee %
For a kind of observers ${O}_H$  at the point $O|_{\xi^\alpha=0}$
with $\alpha$ takes three of
 $1,\cdots,4$, which will become the   spatial origin of a corresponding chart of
  the Beltrami atlas,
  this simultaneity is the same with respect to the
 proper-time
 simultaneity on \HsM.

 The generators of the \dS-algebra $\mathfrak{so}(1,4)$  read:
\be\label{Generator}%
i{{\hat {L}}}_{AB} = \xi_A \frac{\partial}{\partial\xi_B} - \xi_B
\frac{\partial}{\partial\xi_A}, \quad \xi_A=\eta_{AB}\xi^B,
\ee%
which are  proportional to the Killing vectors on the \HsM. They
form an $\mathfrak{so}(1,4)$-algebraic relation and  the
5-dimensional  angular momentum (\ref{angular5}) can also be viewed
as a
set of Noether's charges of the particle 
with respect to these Killing vectors.

The first Cisimir operator of the algebra corresponding to the Einstein-like formula
(\ref{angular5}) is%
\be\label{C15}%
 \hat C_1:= -\frac 1 2 R^{-2} {\hat {L}}_{AB} {\hat {L}}^{AB},\quad
{\hat {L}}^{AB}:=\eta^{AC}\eta^{BD} {\hat {L}}_{CD},%
\ee%
with eigenvalue   $m^2_R$, which gives rise to the classification
of the mass    $m_R$.

\subsubsection{Beltrami model of \dS-spacetime and inertial 
motions}

Let us now consider   the
 \BdS-spacetime and uniform motions along straight wold-lines on it.

\omits{There is a Beltrami atlas   covering the
\BdS-spacetime chart by chart with all 
transition functions on intersections of different charts are of
\FLT-type.}  In order to preserve the orientation, for an
intrinsic geometry of the \BdS-space, it is also needed an atlas
with eight charts   $U_{\pm a}:= \{ \xi\in{H}_+ : \xi^a\gtrless
0\}, a=1, \cdots, 4$ \cite{BdS, BdS2}.

In the charts $U_{\pm 4}$, for instance, the Beltrami coordinates
are
\be \label{u4}%
x^i|_{U_{\pm 4}} =R \frac{\xi^i}{\xi^4},\quad i=0,\cdots, 3;\quad
\xi^4|_{U_{\pm 4}}=({\xi ^0}^2-\sum _{\a=1}^{3}{\xi ^\a}^2+ R^2
)^{1/2} \gtrless 0.
\ee%
In the charts $\{U_{\pm a}, a=1,2,3\}$,
\begin{equation}  %
y^{j'}|_{U_{\pm a}}=R\frac{\xi^{j'}}{\xi^a},\quad
j'=0,\cdots,\hat{a}\cdots,4; \quad \xi^{a}|_{U_{\pm a}}\gtrless 0.
\end{equation}
Then all
transition functions are of \FLT.
\omits{For example, in the intersection $U_{+4}\cap U_{+3}$, the
transition function $T_{+4,+3} =\xi^3/\xi^4=x^3/R=R/y^4$ so that
$x^i=T_{+4,+3}y^{i'} $ for
$i=i'=0,1,2$ and $x^3=R^2/y^4$. }

In the chart $U_{+4},~ \xi^4 > 0$,   the observer ${O_H}|_{\xi^a=0},
(a=1,2,3)$ on the \HsM\ (\ref{H+}) now becomes an observer $O_I$
rest at the spatial origin ($x^a=0$). And there are  domain
condition, Beltrami metric and
boundary condition as follows %
\be\label{domain} %
{B}d{S}:&& \sigma(x):=\sigma(x,x)=1-R^{-2} \eta_{ij}x^i x^j
>0,\\\label{metric}%
&&ds_+^2=[\eta_{ij}\sigma^{-1}(x)+ R^{-2} \eta_{il}\eta_{jk}x^l x^k
\sigma^{-2}(x)]dx^i dx^j,\\\label{bdrbds} %
\partial_P ({B}d{S}): && \sigma(x)=0, \ee%
where   $(\eta_{ij})_{ij=0,\cdots, 3}={diag} (1, -1,-1,-1)$.
 They are invariant
under $FLT$s of $SO(1,4)$ sending a point $A(a^i)$ to the origin
$O(o^i)$ with all
 coordinate $o^i=0$:
\be\nno%
 x^i\rightarrow \tilde{x}^i&=&\pm
\sigma^{1/2}(a)\sigma^{-1}(a,x)(x^j-a^j)D_j^i,\\\label{G}
D_j^i&=&L_j^i+ { R^{-2}}%
\eta_{j l}a^l a^k (\sigma(a)+\sigma^{1/2}(a))^{-1}L_k^i,\\\nno
L&:=&(L_j^i)_{i,j=0,\cdots,3}\in SO(1,3). %
\ee

 \omits{
Thus ${SR}_{c, R}$ offers a consistent way to define a set of
observable for free particles. These issues
significantly confirm  that the motion of a free particle 
 together with the Beltrami coordinate atlas and
corresponding observers  of the system are all of inertia.}

For a free particle with mass  $m_R$, its  uniform `great
circular'
 motion along a geodesic on the \dS-hyperboloid \omits{
having a set of conserved observables  as the 5-dimensional
angular momentum with an Einstein-like formula, }now  becomes a
uniform motion along the time-like geodesic as a straight
world-line on the \BdS-spacetime.
In fact, such a time-like geodesic 
is equivalent to 
\be\label{t-geod}%
\frac{d p^i}{ds_+}=0,\quad{p^i}:=m_R\sigma^{-1}(x)\frac{dx^i}{ds_+}.%
\ee%
Thus, 
${p^i}
={ consts}$. %
And  the coordinate velocity components $v^a=dx^a/dt$
 are constants: %
\be\label{umn}%
\frac{p^a}{p^0}=\frac{dx^a}{dx^0}=:c^{-1}v^a=consts, \quad x^0=ct,\quad a=1,2,3.%
\ee%
It can be integrated further to get  linear result as a counterpart
of  (\ref{sl}).

For  massless particles or light signals, similar issues hold as
long as the proper-time $s_+$ is replaced by an affine parameter
$\lambda_+$.

Under the \FLT s (\ref{G}) of $SO(1,4)$, all these properties 
together with  Beltrami systems are transformed among themselves.
And these properties are well defined chart by chart.

It should be noted that in principle we may also introduce two other
sets of inhomogeneous projective coordinates  without antipodal
identification by $\tilde x^j:=R \xi^j/\xi^0, j=1,\cdots,4;$ or by
$\check x^j_\pm:=R\xi^j/(\xi^0\pm \xi^4)$. However, if we require
that under limit $R\to \infty$ the coordinates and the
transformations among them are back to the \Mink-coordinates and
their Poincar\'e transformations, only the Beltrami atlas with
coordinates $x^j$ in (\ref{u4}) survive.

\subsection{Klein's Erlangen program versus  principle of
relativity in all possible kinematics}

As was emphasized, in analogy with that weakening  Euclid's fifth
axiom leads to Riemann and Lobachevsky geometries on an almost equal
footing with Euclid geometry,  there should be  two other kinds of
\dS/\AdS\ \SR\
on an almost equal footing with Einstein's one.

In fact, there is a one-to-one correspondence between  these
geometries  on maximally symmetric spaces $\mathfrak{S}_E$ with
maximum symmetries $\mathfrak{G}_E$ and their physical
counterparts  on maximally symmetric spacetimes $\mathfrak{S}$
with maximum symmetries $\mathfrak{G}$.
We list them in the following table:%

\bc
\begin{tabular}{lcl}
\quad Table 1.  Correspondence between & &\hskip -12mm 4-d geometry and 3+1-d special relativity \\
\hline { \quad Geometry  on $\mathfrak{S}_E$ with
$\mathfrak{G}_E$} &\qquad  ~~&\qquad
{ Spacetime Physics  on $\mathfrak{S}$ with $\mathfrak{G}$}\\
\hline \\[-5mm]
 ${E}^4/{S}^4/{L}^4$ as $\mathfrak{S}_E$  &  &
\qquad ${M}^{1,3}$/\dS$^{1,3}$/\AdS$^{1,3}$
 as $\mathfrak{S}$\\
$ISO(4)/SO(5)/SO(1,4)$ as $\mathfrak{G}_E$& & \qquad $ISO(1,3)/SO(1,4)/SO(2,3)$  as $\mathfrak{G}$\\
Descartes/Beltrami systems&  &\qquad  Minkowski/Beltrami systems\\
Points & 　 & 　\qquad Events\\
Geodesics as straight lines&  & \qquad Geodesics as straight world-lines \\
Principle of Invariance & & \qquad  Principle of Relativity\\
Erlangen Programm\qquad  &  &  \qquad Theory of Special Relativity\\
\hline \\
\end{tabular}

 \ec

From the  viewpoint of   \dS/\AdS\ \SR, all  possible kinematics 
can be set up based on the corresponding \PoR\ and the
corresponding postulate of universal constant(s), respectively,
although for Newton's theory these constants are all degenerate
(see, e.g., \cite{NH}).

Actually,  in  view of  geometrical and algebraic contractions,
there are some important contraction relations among these
kinematics. Namely, all other kinds of kinematics can be viewed as
some contraction of the \dS/\AdS\ \SR\ under certain limit of the
constant(s): Einstein's \SR\ on  \Mink-spacetime with Poincar\'e
\PoR\ of Poincar\'e invariance under $R\to \infty$; Newton's
mechanics on Newton's space and time with Galilei \PoR\ of Galilei
invariance under $R, c, \to \infty$. Newton-Hooke/anti-Newton-Hooke
mechanics on Newton-Hooke/anti-Newton-Hooke space-time with
Newton-Hooke \PoR\ of Newton-Hooke/anti-Newton-Hooke symmetry under
the Newton-Hooke limit: $R, c, \to \infty$, but the Newton-Hooke
constant $\nu:=c/R=const$, respectively \cite{NH}.

Conversely, there are also some `deformation' relations among them.

\omits{Therefore, the role of the \PoR\ on \dS/\AdS-spacetime
proposed by Lu for physics  is in analogy with that of Klein's
Erlangen programm for
geometry. 
}

\subsection{Historical remarks}

It should be noted that the \dS\ geometry and physics are studied
for long time in the framework of general relativity.
However, 
the \PoI, the law of inertia and relevant physics on \dS\
spacetime had been missed, although   Beltrami systems had been
used or mentioned time after time  in literatures.

As early as in 1917,  de Sitter \cite{dS17} introduced
Beltrami-Kliein coordinates for his solution in the debate with
Einstein on `relative inertia'.  Their debate also drew attentions
from Klein and Weyl. A few years later, Pauli mentioned fractional
linear transformations and the Beltrami model of 4-dimensional
Riemann sphere in his famous book but ignored their possible
physical applications \cite{Pauli20}. Snyder \cite{Snyder47}
proposed a quantized space-time model in projective geometry
approach, explained by Pauli,  to \dS-space of momenta. This is in
fact the earliest and simplest model among the `doubly special
relativity' or the `deformed spacial relativity'  widely studied
recently \cite{dsr}. Although there is a simple one-to-one
correspondence between Snyder's model and the \dS\ \SR\
\cite{duality}, it had not been  considered what should be the
counterpart in coordinate picture of Snyder's model in momentum
space before. Schr\"odinger also proposed the `elliptic explanation'
of \dS-spacetime concerning the antipodal identification
\cite{Sch56}, which has  been also studied in \cite{dSZ2}. However,
there had been no study on such a key issue for long time that  in
either Beltrami coordinates or inhomogeneous projective ones there
are uniform motions along time-like or null geodesics. Therefore,
there should be {\it the law of inertia} on  \dS-spacetime and these
coordinates should play a role of {\it inertia}.

On the other hand, Umov, Weyl and Fock (see, e.g., \cite{Fock})
studied the \FLT s as most general transformations among inertial
systems and inertial motions. But, they  did not relate these \FLT
s to either  Beltrami systems  or the inertial motions on them.
Otherwise, the inertial law on \dS/\AdS-spacetimes could be
discovered long time ago.

Since  1950s,  Hua and Lu develop the theory of classical domains
and harmonic analysis on the domains \cite{HuaLu}. As the
 Beltrami model of hyperboloid  is a special case, Hua and Lu
use the generalized Beltrami metric widely in their studies. In
1970, Lu \cite{Lu} first noticed the key point in physics and began
the research on the \dS/\AdS\ \SR\ later \cite{Lu, LZG}. Promoted by
recent observations on the dark universe, further studies are made
\cite{lu05, BdS, BdS2, BdS3, IWR, TdS, NH, yan, duality, OoI, PoI}.

%
%

\section{Principle of Relativity  and  De Sitter Special  Relativity}\label{dSSR}

We now briefly introduce the properties of the \dS\ \SR\ based on
the \PoR\ and the postulate on invariant universal constants. We
 show its cosmological significance via the coin-like model of \dS-space with both
 the \PoR\ and the \CP. We also  explain why  our universe should be slightly closed
 if it is asymptotic to a \dS-space and why its evolution
can fix on a kind of Beltrami inertial frames together with all its
contractions. Thus,  our universe
displays as the origin of inertia without Einstein's `argument in a circle' for the \PoI.

\subsection{Transformations among inertial systems and  principle of relativity}

The existence of the  \dS/\AdS\ \SR\  can also be prospected from
another angle: What are the most general transformations among
inertial motions and inertial systems? As was just mentioned, Umov,
Weyl and Fock \cite{Fock} studied this problem long time ago.

As in both Newton's mechanics and Einstein's \SR,   inertial
motions can be defined as a kind of motions with  uniform
coordinate velocity along  straight lines in a kind of coordinate
systems. Namely, if in a system $S(x)$ for a free particle its
motion  satisfies
\be\label{uvm}%
 x^\alpha=x_0^\alpha+v^\alpha(t-t_0),\quad
v^\alpha=\frac{dx^\alpha}{dt}=consts,\quad \alpha=1, 2, 3, %
\ee
the motion  and the system are called  inertial one, respectively.

\omits{Thus we may start with eqn. and to get \omits{what Fock
proved to get Fock's theorem as a lemma denoted as }the lemma 1.}
\omits{Fock proved a theorem on the general transformations among
uniform velocity motions \cite{Fock}.}

Let us consider a transformed  system  $S'$, if the same particle
is described by
\be\label{uvm'}%
{x'}^\alpha={x'}_0^\alpha+{v'}^\alpha(t'-t'_0), \quad
 {v'}^\alpha=\frac{d{x'}^\alpha}{dt'}=consts,
 \ee
the transformed system is also of inertia. What are the most general transformations between these two inertial systems? 
Fock \cite{Fock} showed that  the most general form of  transformations 
\be\label{FockT}%
{x'}^i=f^i(t, x^\alpha),\quad i=0, \cdots, 3, %
\ee
which transform a uniform straight line motion in $S$ with
(\ref{uvm}) to a motion of the same nature in $S'$ with
(\ref{uvm'}) should be that  four functions $f^i$ are ratios of
linear functions, all with the same denominator. Thus, they are of
the $FLT$-type.

As was mentioned, in general 
we may not assume that the proper-length of a
 `rigid' ruler and the  proper-time of an `ideal' clock be Euclidean.
 In other words,
 the spatial coordinates themselves and the temporal coordinate itself are not
 assumed to be uniform in the Euclidean sense, respectively.
 This is different from either Newton's mechanics or
Einstein's \SR. Otherwise, the \FLT s should just be the linear ones
in Newton's mechanics or Einstein's \SR.  Fock just did so by
assuming the wave front equation with \Mink-metric so that the
 \FLT s  reduce to the transformations of Poincar\'e group.

As there is a \Mink-metric on  4-dimensional \Mink-spacetime
invariant under transformations of Poincar\'e group with ten
parameters, we should require that there be a metric in the
inertial systems on 4-dimensional spacetime
 and the $FLT$s form a group with ten parameters, like  Galilei
 group in Newton's mechanics and  Poincar\'e group in Einstein's
 \SR , including four for spacetime `translations', three for boosts,  and
 rest three for space
 rotations.
 Thus, according to the properties of maximally symmetric spaces (see, e.g.,
  \cite{weinberg}), such kind of 4-dimensional  spacetimes
  should be the maximally
symmetric spacetimes $\mathfrak{S}$ of positive/negative constant
curvature with radius $R$ or zero curvature with $R \rightarrow
\infty$. Namely, they are just the
 \dS/\AdS/\Mink-spacetime being the maximally symmetric
spacetime $\mathfrak{S}$ with
$SO(1,4)/SO(2,3)/ISO(1,3)$-invariance being the maximum symmetry
$\mathfrak{G}$, respectively.

From the  viewpoint of  projective transformations, these are
obvious: those uniform motions along straight lines are of
projective and the transformations of projectively \FLT s. All the
maximally symmetric spacetimes with maximum symmetries,
respectively, are of sub-geometries of projective geometry. This is
also in consistency with Klein's program \cite{klein1872}. Of
course, the orientation should be preserved in physics.


As was mentioned, for the \dS/\AdS-spacetime  Beltrami systems
are indeed these systems and  the observer $O_I$ at
the spatial origin is of inertia. 
Therefore, on the \BdS/anti-\BdS-spacetime, there  are  also the
\PoR\  and the postulate on invariant universal constants. The \PoR\
states: {\it The physical laws without gravity are invariant under
the group transformations among inertial systems on the
4-dimensional \dS/\AdS-spacetime, respectively.} The postulate
requires: {\it In the inertial systems on 4-dimensional
\dS/\AdS-spacetimes, there are two invariant universal constants,
the speed of light $c$ and the curvature radius $R$}.

Based on the principle and the postulate, the \dS/\AdS\  \SR\ can be
set up \cite{BdS, BdS2, BdS3}.

\subsection{Law of inertia, generalized Einstein formula, light cone and horizon}

Thus, there is a Beltrami atlas of inertia on the \BdS\ and in each
chart there are  condition (\ref{domain}),  metric (\ref{metric})
and  \FLT s (\ref{G}) of \dS-group.

 In such a \BdS, the generators of $FLT$s   read
\be\nno
  {\hat p}_i &=&(\delta_i^j-R^{-2}x_i x^j) \partial_j,~~
  x_i:=\eta_{ij}x^j,\\\label{generator}
  {\hat L}_{ij} &=& x_i {\hat p}_j - x_j {\hat p}_i
  = x_i \partial_j - x_j \partial_i \in \mathfrak{so}(1,3), 
\ee
and form an $\mathfrak{so}(1,4)$  algebra %
\be\nno%
  [ \hat{p}_i, \hat{p}_j ] &=& R^{-2} \hat{L}_{ij},~~  
  {[} \hat{L}_{ij},\hat{p}_k {]} =
    \eta_{jk} \hat{p}_i - \eta_{ik} \hat{p}_j,
\label{so14}\\
  {[} \hat{L}_{ij},\hat{L}_{kl} {]} &=&
    \eta_{jk} \hat{L}_{il}
  - \eta_{jl} \hat{L}_{ik}
  + \eta_{il} \hat{L}_{jk}
  - \eta_{ik} \hat{L}_{jl}. 
\end{eqnarray}

 For a free  particle  along a time-like geodesics being a straight world-line
  there is a set of 
 conserved
quantities $p^i$ in (\ref{t-geod}) and 
 \be\label{angular4}%
 L^{ij}=x^ip^j-x^jp^i,\quad && \frac{dL^{ij}}{ds_+}=0.
\ee%
These are 
 pseudo 4-momentum $p^i$, pseudo 4-angular-momentum $L^{ij}$ of the
particle, which  constitute the conserved 5-dimensional angular
momentum as was shown in (\ref{angular5}).

Thus, there is a law of inertia on \dS/\AdS: {\it The free
particles and light signals without undergoing any unbalanced
forces should keep their uniform motions along straight
world-lines in  linear forms in  Beltrami systems  on
\dS/\AdS-space, respectively.}

The equation  of motion for a forced particle can also be given
\cite{BdS2, BdS3}.

Further, all these conserved quantities  satisfy a generalized
Einstein
formula  on \BdS-space from the Einstein-like formula (\ref{5eml}): %
\begin{eqnarray}\label{eml}%
 E^2=m_{R}^2c^4+{p}^2c^2 + \d
{c^2} {R^2} { j}^2 - \d {c^4}{R^2}{
k}^2,%
\end{eqnarray}
with energy $E=p^0$, momentum $p^\alpha$,
$p_\alpha=\delta_{\alpha\beta}p^\beta$, `boost' $k^\alpha$,
$k_\alpha=\delta_{\alpha\beta}k^\beta$ and 3-angular momentum
$j^\alpha$, $j_\alpha=\delta_{\alpha\beta} j^\beta$.  And these
observables may also be viewed as Noether's charges of the particle
with respect to the Killing vectors proportional to the generators
in (\ref{generator}). Note that $m^2_{R}$ now is the eigenvalue of
first Casimir operator of \dS-group, the same as the one in
(\ref{C15}).

If we introduce the Newton-Hooke constant $\nu$
\cite{NH} and link the  radius $R$ with the cosmological constant $R\simeq (3/\La)^{1/2}$,%
\be\label{NHc}%
\nu:=\d {c} {R}\simeq c (3/\La)^{-1/2},\quad \nu^2 \sim 10^{-35}s^{-2}. %
\ee%
It is so tiny that all experiments that prove Einstein's special
relativity at ordinary scales cannot exclude  the \dS\ \SR. However,
from the algebraic relation (\ref{so14}) and this important formula,
it  qualitatively follows that for all celestial objects including
the CMB as test objects in the cosmic scale, their free motions are
always with both the conserved energy-momentum and the angular
momentum.

The interval between two events and   light-cone can be well
defined as the inverse Wick rotation counterparts of (\ref{AB})
and (\ref{ABL}), respectively. In fact, for two separate events
$A(a^i)$ and $X(x^i)$ 
\be\label{lcone0} %
 \Delta^2_R(a, x) = R^2\,[\sigma^2(a,x)-\sigma(a)\sigma(x)]%
\ee %
is invariant under the \FLT s of $SO(1,4)$.
 Thus, the interval between $A$ and
$B$ is time-like, null or space-like,
respectively, according to%
\begin{equation}\label{lcone}%
\Delta_R^2(a, b)\gtreqless 0.%
\end{equation}
The proper length of time/space-like 
interval between $A$ and $B$ is the integral of line element $ds_+$
in (\ref{metric}) over the geodesic segment $\overline{AB}$:
\be \label{AB1}%
S_{t-like}(a, b)&=&R \sinh^{-1} (|\Delta(a,b)_R|/R), \\
\label{AB1sl} S_{s-like}(a,b)&=& R \arcsin (|\Delta(a,b)_R|/R).%
\ee
\omits{where ${I}=1, -i$ for timelike or spacelike, respectively.}

The Beltrami light-cone at an event $A$ with running events $X$ is
\be \label{nullcone} %
{F}_{R}:= R
\{\sigma(a,x) - [\sigma(a)\sigma(x)]^{1/2}\}=0.%
 \ee%
It satisfies the null-hypersurface condition.\omits{
 \be\label{Heqn}%
\left . g^{ij}\frac{\partial f}{\partial x^i}\frac{\partial
f}{\partial x^j}\right |_{f=0}=0, %
\ee
where $g^{ij}=\sigma(x)(\eta^{ij}-R^{-2} x^i x^j)$ inverse of
(\ref{bhl}).} %
At the origin %
$a^i=0$, the light cone becomes a \Mink-one $\eta_{ij}x^i x^j=0$ and 
$c$ is numerically the velocity of light in the vacuum.

There is also a horizon tangent to the boundary on \BdS\ for the
observers ${O}_I$:
\be\label{Heq}%
\lim_{a \to a'} \sigma (a,x)=0, \qquad \lim_{a\to a'}\sigma(a)=0.%
\ee%

For the horizon in  Beltrami systems, it  is actually at $T=0$
without entropy. But,  at the horizon in other \dS-spacetimes,
such as the static \dS-universe and the Robertson-Walker-like
\dS-spacetime, Hawking temperature and area entropy appear as
non-inertial effects rather than gravitational ones \cite{TdS}.
Thus, \dS-spacetime is completely different from  black hole.

\subsection{Two kinds of  simultaneity,   principle of relativity and cosmological principle}

In order to make  measurements, simultaneity should be defined. As
was mentioned, different from Einstein's special relativity, there
are two kinds of simultaneity related to two kinds of measurements
with respect  to the \PoR\   and the cosmological principle,
respectively. It is important that these two kinds of simultaneity
together with corresponding principle are very closely related to
each other just like a coin with two sides.

In the contraction $R\to \infty$, however, they coincide with each
other.

\subsubsection{Beltrami simultaneity for \PoR} 

 Let us first consider the Beltrami simultaneity with respect to the
 Beltrami time coordinate.
 For an inertial observer ${O}_I$ at the spatial origin of the system, who is just the observer
  $O_H$ for the uniform `great
 circular' motion (\ref{angular5}) on the  \HsM, two events ($A, B$)
 are simultaneous if and only if their Beltrami time coordinates are equal to each other 
\be %
a^0:=x^0(A) =x^0(B)=:b^0. 
\ee%
This simultaneity defines a $3+1$ decomposition of the \BdS-matric
(\ref{metric})
\be%
 ds^2 =  N^2 (dx^0)^2 - h_{\a\b} \left (dx^\a+N^\a dx^0 \right
)
\left (dx^\b+N^\b dx^0 \right ) , \quad \a,\b=1,2,3,%
\ee
with lapse function, shift vector and induced 3-geometry on
3-hypersurface $\Si_c$ in one coordinate chart, respectively
%
\begin{eqnarray}\nno
N\ &=&\{\si_{\Si_c}(x)[1-(x^0 /R)^2]\}^{-1/2}, \\%
N^\a&=&x^0 x^\a[ R^2-(x^0)^2]^{-1},  \\\nno h_{\a\b}&=&\dl_{\a\b}
\si_{\Si_c}^{-1}(x)-{ [R\si_{\Si_c}(x)]^{-2} \dl_{\a\gamma}
\dl_{\b\dl}}x^\gamma
x^\dl ,\\\nonumber%
 &&\si_{\Si_c}(x)=1-(x^0{/R})^2 + {\dl_{\a\b}x^\a x^\b
/R^2}.\nno
\end{eqnarray}
It is easy to see that at $x^0=0$, $\si_{\Si_c}(x)=1+{\dl_{\a\b}
x^\a x^\b/R^2}, ~N=\si_{\Si_c}^{-1/2}(x), ~N^\a=0.$ \omits{Then the
Cauchy hypersurface is $ \Si_c \simeq S^3$.
 And at  Beltrami time $x^0\neq 0$, as long as $x^0$ is still time-like,
 we should also have {$ \Si_c \simeq S^3$}.}

This simultaneity leads to a definition of non-Euclidean Beltrami
ruler and its relation to  spatial coordinate distance of two
simultaneous events. A Beltrami
 ruler
at  $x^0$ is defined by 
\be%
dl_B^2|_{x^0}=-h_{\a\b}|_{x^0}dx^\a dx^\b.%
\ee

 In fact, all measurements in the
 Beltrami systems
  are in analogy with that on  \Mink-spacetime
 as long as it is are no longer Euclidean not only the Beltrami time and proper-time
 of a standard clock as well as their relation, but also the Beltrami spatial
 coordinates and the proper-length of a ruler as well as their relation.

\subsubsection{Proper-time simultaneity  and  Robertson-Walker-like 
 \dS-space}

 Another simultaneity is the same as the one in (\ref{simtH}) for the
 observer $O_H$.

The proper-time $\tau$ of a clock rest  at spatial origin $x^a=0$ of
Beltrami
 system
relates the coordinate time $x^0$ 
as
\begin{eqnarray}\label{ptime}%
 \tau=R \sinh^{-1} (R^{-1}\sigma^{-\frac{1}{2}}(x)x^0)+ \tau_0,
\end{eqnarray}
where $\tau_0$ is a constant to be determined by physical consideration. For the sake of
simplicity,
we may simply take $\tau_0=0$. With respect to this proper-time $\tau$\omits{ of a clock rest at
the spatial origin of the Beltrami systems}, the proper-time
simultaneity can be defined as: The events are simultaneous if and
only if their proper time $\tau$ is the same
\begin{equation}\label{smlt}
 x^0\sigma^{-1/2}(x,x)=(\xi^0:=)R \sinh(\tau/R)=\rm const.
\end{equation}
 If $\tau$ is taken as a  temporal coordinate together with the spatial Beltrami coordinates,
 the \BdS-space becomes
a Robertson-Walker-like \dS-model with a metric having $\tau$ being a `cosmic'-time:%
\be\nno%
 ds^2&=&d\tau^2-dl_C^2=d\tau^2- \cosh^2(\tau/R) dl_{0}^2,\\\label{RWds}%
dl_{0}^2&=& {\{\delta_{\a\b}\sigma_{\Sigma_\tau}^{-1}(x)
-[R\sigma_{\Sigma_\tau}(x)]^{-2}\delta_{\a\gamma}\delta_{\b\dl}x^\gamma
x^\dl\}}
 dx^\a dx^\b, \\\nno 
&& \sigma_{\Sigma_\tau}(x,x)=1+R^{-2}\delta_{\a\b}x^\a x^\b
>0,
\ee
%
where $dl_{\omits{\Sigma_\tau} 0}^2$  a 3-dimensional Beltrami
metric on an $S^3$ of radius $R$. This is  an `empty' cosmic model
with an accelerated expanding and slightly closed cosmos of
curvature in the order of $O(R^{-2})$.

Thermodynamically, from Eq. (\ref{ptime}), it is easy to see that
for  the proper-time, there is a period in the imaginary proper-time
that is inversely proportional to the Hawking-temperature
${c\hbar/(2\pi R k_B)}$ at the horizon. If the temperature Green's
function can still be applied here, this should indicate that  there
are Hawking-temperature and `area' entropy $S= \pi R^2
{c^3k_B/G\hbar}$ at the horizon in the Robertson-Walker-like
\dS-space (\ref{RWds}). But, they are not caused by gravity rather
by non-inertial motions. This is also in analogy with relation
between Einstein's special relativity in \Mink--space and the and
the horizon in Rindler-coordinates. The temperature at the
Rindler-horizon is caused by non-inertial motion rather than gravity
\cite{TdS}.


Since there is a relation between two kinds of simultaneity for the
\PoR\ and the cosmological principle, \dS-spacetime provides a coin-like model
for these two principles.

On one side, with Hawking temperature and `area' entropy there is
the Robertson-Walker-like \dS-cosmos with cosmological constant
fitting the \CP. And on another side, at zero temperature without
 entropy  there is the \BdS-spacetime
with the \PoI.  Thus, the former should just display as the origin
of law of inertia on the
latter and the principle of inertia on the latter 
provides a benchmark for physics on  \dS-space including both the
\BdS-space and the Robertson-Walker-like \dS-cosmos.

In other words, 
on  \dS-spacetime there is a kind of {\it inertial-comoving-like
observers}, ${O}_{I-C}$, equipped a type of two-time-scale timers 
of  Beltrami time and  `cosmic'-time, as well as  corresponding
rulers. They may act as inertial observers $O_I$ or comoving-like
ones $O_C$  in different experiments or observations,
respectively, reflecting these principles and their important
relation. \omits{on \dS-invariance. Their different roles listed
in the following table.

\bc  \begin{tabular}{lcl} {Inertial  Observers ${O}_{I}$}  \quad &{
$\rightleftarrows$}
&\quad {Comoving-like  Observers ${O}_{C}$}\\
Experiments in lab. \quad & 　&\quad Cosmic observations\\
 Beltrami model \quad & 　&\quad Robertson-Walker-like \dS-model\\
Beltrami timer \quad & 　&\quad  `Cosmic'-time timer\\
Beltrami ruler \quad & 　&\quad  Comoving-like ruler\\
$\qquad  \cdots $ \quad & 　&\quad  $\qquad\cdots$
\end{tabular}
\ec
}Actually, once the observers would carry on  experiments in their
laboratories, they should  switch on  Beltrami time and off
`cosmic'-time  so that they act as inertial observers ${O}_{I}$
and all observations are of inertia. When they would take
approximatively `cosmic' observations on  distant stars and
 cosmic objects other than the cosmological constant as test objects
 they should  switch off
Beltrami time and on  `cosmic'-time  again, so they should act as
a kind of comoving observers ${O}_{C}$ as they hope. Namely, what
should be done for those  inertial-comoving-like observers
${O}_{I-C}$ is
just to switch   off  `cosmic'-time and on 
 Beltrami time once they want to be back to local experiments
from their comoving-like 
observations 
and vice versa.

It is worth while to mention that in general there are  other kinds
of \dS-comoving coordinates with  flat or open 3-dimensional cosmos,
respectively. However, from the viewpoint of \dS\ \SR, the above Robertson-Walker-like
\dS-comoving coordinates in (\ref{RWds}) with closed 3-dimensional cosmos is most natural and simplest
among all of them.

\subsection{Cosmological significance of de Sitter special relativity}
\omits{ proper-time simultaneity  and the Robertson-Walker-like 
 coordinates}

If our universe is accelerated expanding and possibly asymptotic to
a \dS, its fate should be  a Robertson-Walker-like \dS-space. This
is very natural from the  viewpoint of   \dS\ \SR. Thus, there is remarkable
cosmological significance for \dS\ \SR\ different from the conventional approach in \GR.

First, there is an important prediction. 

If our universe is asymptotic to  the Robertson-Walker-like
\dS-space (\ref{RWds}) of $R^2\simeq3\Lambda^{-1}$ with `area'
entropy, the 3-dimensional cosmic space of the dark universe should
be closed and asymptotic to an accelerated expanding $S^3$ with an
entropy bound $S\simeq3\pi {c^3k_B/\Lambda G\hbar}$. Its deviation
from the flatness is in the order of the cosmological constant
$O(\Lambda)$.

This is in consistency with recent data from WMAP \cite{WMAP} and can be further checked. %

On the other hand, the evolution of our unverse can determine the Beltrami
inertial frames of the \PoI\ in \dS\ \SR\ and all other
kinds of inertial frames contracted from the Beltrami frames.

As is well known, according to Einstein, there is an `argument in a circle' for the
\PoI. In his most famous book, Einstein wrote: 
 `The weakness of the principle of inertia  lies in this, that it
 involves an argument in a circle: a mass moves without acceleration
 if it is sufficiently far from other bodies; we know that it is
 sufficiently far from other bodies only by the fact that it moves
 without acceleration. Are there at all any inertial systems for
 very extended portions of the space-time continuum, or, indeed, for
 the whole universe? We may look upon
 the principle of
 inertia as  established, to a high degree of approximation, for the space of
  our
 planetary system, provided that we neglect the perturbations due to the sun and
 planets. Stated more exactly, there are finite regions, where, with respect to
  a suitably
 chosen space of reference, material particles move freely without
 acceleration, and in which the laws of the special theory of
 relativity, $\cdots$, hold with remarkable accuracy. Such regions
 we  shall call ``Galilean regions".' \cite{1923}

 `Are there at all any inertial systems $\cdots$ for
 the whole universe?'  Einstein  raised such a severe question, but he did not answer.

 With the help of asymptotic behavior of our universe and the \dS\ \SR, this question
 can be  definitely
 answered.  In fact, for the \PoI\ on \dS-spacetime,  there is
 no Einstein's  `argument in a circle' and  the inertial frames of \BdS-type
 do exist for the whole
 universe.\omits{, since   the 
 evolution of our
 universe 
 can determine  them.} 
Actually, without measuring any acceleration of a mass, all  needed are  the  time arrow and approximative
symmetry of our universe roughly described by the
cosmological principle. %

 If our
universe is asymptotic to the Robertson-Walker-like \dS-space
(\ref{RWds}), the time arrow and the homogeneous space of our
universe should coincide with the `cosmic'-time arrow and tend to 
an accelerated expanding $S^3$ of the Robertson-Walker-like
\dS-space, respectively. These pick up the directions of the `cosmic' temporal
axis and the spatial axes for the Robertson-Walker-like
\dS-systems up to  spatial rotations of  $SO(4)$  among all them
related by \dS-transformations so that the \dS-symmetry reduces to
its subgroup $SO(4)$ of the Robertson-Walker-like \dS-cosmos with
a `cosmic'-time, the direction of which coincides  with  the time
arrow of our universe. Then, via the important relation
 between  Beltrami systems  and the Robertson-Walker-like \dS-model, i.e.
 via the relation
(\ref{ptime}) between the `cosmic'-time  and the Beltrami time,
 the directions of the axes in a kind of  Beltrami frames  can be given.
 This is just like to flip a coin
from one side to another. In fact, the Beltrami temporal axis is
related to the axis of `cosmic'-time on the Robertson-Walker-like
\dS-space and the spatial axes of the Robertson-Walker-like
\dS-space (\ref{RWds}) are just the Beltrami spatial coordinates.
Thus, the evolution of our universe can fix on this kind of
Beltrami frames  in such a way that there is no Einstein's `argument in
a circle', since
 gravitational effects and acceleration of a mass   do
  not explicitly play any roles here.

 There are two invariant universal constants, $c$ and $R$, in the Beltrami frames.
In order to set up the real Beltrami frames, it is also needed to
determine their value numerically. If so, how can present
experiments or observations nowadays determine their values in the
fate of our universe? How in these present experiments or
observations we can neglect the gravitational effects?

In fact, although the  Beltrami frames of inertia   depend on
the dimensions of these two invariant universal constants\omits{: speed
as dimension of $c$ and length as dimension of $R$}, the property
of inertia for the frames does not depend on their concrete
values unless for measurements of concrete physical processes. Of
course, \omits{in order to make the
measurements/observations of concrete physical processes in our
universe}physically, their concrete values are certainly needed and should be
determined by two kinds of experiments or observations. Since
these constants are supposed to be invariant and universal
approximately, the value of  $c$  should still be taken
as the one in Einstein's \SR. Note that this also fixes on
 the origin of  Beltrami frames  since
 the light cone (\ref{nullcone}) at the origin is just Minkowskian  at present approximatively.
 As for  the value of $R$, it  may
also be taken as $R\simeq(3/\Lambda)^{1/2}$, where the
cosmological constant $\Lambda$ is given by the precise cosmology.
\omits{ based on general relativity as a framework for the data
analysis. }Although the determination of 
$\Lambda$ may depend on some gravitational effects nowadays and so
does the value of $R$, this does not a matter in principle
for fixing on the inertial systems. In fact, changing the value of
$R$ may lead to the conformal extension of  \dS-spacetime, which
will be explained later.

Since in all possible kinematics based on  principle of inertia the
inertial  frames can be given under certain contracting limit
from the Beltrami frames, respectively,  all  different kinds
of inertial frames in the kinematics should also be fixed on by the
evolution our universe  without Einstein's `argument in a
circle' so long as they are regarded as successors of the Beltrami
systems. However, if it is ignored this successive relation of the
inertial systems, the  coin-like relation between the
\PoR\  and the cosmological principle should no longer appear or becomes
trivial in Einstein's \SR\ and Newton's mechanics, except the
Newton-Hooke one.

In addition, if it is further required that in the spacetimes with
gravity there should exist locally the \PoR\ everywhere and anytime
and the values of $c$ and $R$ should be the same as in the \dS\,
\SR, such kind of local inertial frames with the origin at present can also be fixed on in the
same manner by the evolution of our universe.

Thus,  Beltrami systems  of inertia and their localized version \omits{with the origin being
impliedly taken as at present approximatively }together with their
contracting forms do exist in 
the whole
universe. \omits{Since they as open sets mathematically, they do exist
for  universe. There is nothing with acceleration
of a mass and this is independent of
 gravity in the universe. Therefore, for the \PoI\,
there is no longer Einstein's `argument in a circle'. }In the
sense that these systems can be fixed on by the evolution of our
universe, the universe also plays a role as the origin of inertia
in all these kinematics.

 It should be noted that the Beltrami inertial frames determined by the evolution of our
universe are  a kind of `preferred' frames in the sense that  their temporal
axis is related to   the time arrow of our universe. These
`preferred' inertial frames  still exist under different
contractions. However, this `preference'  does not break the \PoR\
that is for physical laws. In fact, the `preference' only plays certain role
when some comoving-like observations are taken, since its temporal
axis is just transformed from the `cosmic'-time axis of the
Robertson-Walker-like \dS-space that coincides with the  time
arrow of our universe.

This is also true for the local Beltrami frames and all their contractions.
Actually, even in \GR\, once the cosmic observations or background should be taken
into account such  kind of local inertial frames should be taken that
their time axis should coincide with the comoving time axis.
In this sense, this kind of local inertial frames is
 `preferred'. \omits{ among all local inertial frames. display
 although the
Poincar\'e invariant theories are good enough for the ordinary
scale physics,  as
long as . This is
just the case in \GR, but it is hard to explain.}But, in \GR, the symmetry for
 local inertial frames is not the same as that in Einstein's \SR.

\omits{Right after his statement on the `argument in a circle' as the
weakness of the \PoI\  in both Newton's mechanics and his \SR,
Einstein raised a severe question: `Are there at all any inertial
systems for very extended portions of the space-time continuum, or,
indeed, for the whole universe? '\cite{1923} in  view of Einstein,
these inertial systems can exist in what he called `Galilean
regions' where his \SR\ holds `with remarkable accuracy': `We may
look upon the principle of inertia as established, to a high degree
of approximation, for the space of our planetary system, provided
that we neglect the perturbations due to the sun and planets. Stated
more exactly, there are finite regions, where, with respect to a
suitably chosen space of reference, material particles move freely
without acceleration, and in which the laws of the special theory of
relativity, $\cdots$, hold with remarkable accuracy. Such regions we
shall call ``Galilean regions."'\cite{1923} Although Einstein did
not mention explicitly whether there should be such an inertial
system for the whole universe, his answer seems to be negative.

In  view of  the \dS\ \SR, however, the answer is affirmative. {\it
There is a kind of inertial systems on the \BdS-spacetime for the
whole universe, which should be fixed by our
universe and be 
reached in its faraway future. And other kinds of inertial systems
do also exist as the contractions of  Beltrami systems  in
different limits.}

As for Einstein's `Galilean regions', there is another story for
gravity.}

%
%
\section{Conformal Extensions of Three Kinds of Special Relativity
}\label{C3}

 We now consider  conformal
extensions of three kinds of \SR\  and null physics on them as well
as their relations via Weyl
conformal mappings \cite{c3}. 

As is well known, in Einstein's  special relativity on
\Mink-space, massless particles and  light signals move {\it
  in inertia} along null geodesics satisfying  $ds_M^2=0$ invariant under
  conformal group transformations with fifteen parameters.  Thus, symmetry of their
   motions should be enlarged from Poincar\'e group $ISO(1,3)$ to  conformal
  group.
  In the \dS/\AdS\ \SR,  massless particles and light signals move
  also {\it in inertia} along straight lines at
  constant coordinate velocities. Similarly, they also satisfy $ds_{\pm}^2=0$, where
  $ds_+$ is given by (\ref{metric}), invariant
  under conformal group transformations as well. Thus,  symmetry of their
  motions
  should also be enlarged from  \dS/\AdS-group $SO(1,4)/SO(2,3)$ to  conformal
  group.

  In all these cases,
 the conformal extensions  can be realized on a
  null cone $\cal N$ modulo  projective equivalence 
  in a $(4+2)$-dimensional  \Mink\
  space, $[{\cal N}]:=\mathcal{N}/\!\!\sim$ $\subset M^{2,4}$, invariant under the
  conformal group  $SO(2,4)/\mathbb{Z}_2$ with isometry
  subgroup  $ISO(1,3)/SO(1,4)/SO(2,3)$, respectively.
   Further, the null physics
  on  \dS/\AdS/\Mink-spaces can be mapped from one to another by Weyl
  conformal mappings.  In this sense,  there should be a
  triality of these conformal issues \cite{c3}.

  Since the projective
  boundary  of a 5-dimensional \AdS-space, $\partial(AdS^5)$, is just $[\mathcal{N}]$,
   4-dimensional conformal \dS/\AdS-spaces can also be included in $\partial(AdS^5)$,
  in addition to
   conformal \Mink-space.  Thus, if the
  \AdS/\CFT\ correspondence \cite{adscft} is conjectured, there should
  be three versions of \AdS/\CFT\ correspondence \cite{c3}. Further,
  there should be a \dS-spacetime
on the boundary of $S^5 \times AdS^5$ as a vacuum of supergravity.

\subsection{Conformal extensions of \Mink/\dS/\AdS-spaces on a null cone}

 Let us view the \dS/\AdS-space with  radius $R$ as a 
  4-dimensional  hyperboloid ${H}_{R\ \theta}$
  $(\theta = \pm 1 )$ (or simply ${H}_{\theta}$)  embedded in $M^{1,4}$/$M^{2,3}$, respectively: %
  \be\label{5sphr}%
    H_{ \theta}:  &&\eta_{ij}\ \xi^i \xi^j - \theta\,(\xi^4)^2
    = \eta_{\theta AB}\xi^A \xi^B=-\theta R^2 \lessgtr 0\\%
    &&ds^2_{H_\theta}  
    =\eta_{\theta AB}d\xi^A d\xi^B,~~A,B=0,\cdots, 4. %
  \label{ds2pm}\omits{\\\label{bdypm}%
  \partial_P(H_{R \theta}): && \eta_{\theta AB}\xi^A \xi^B=0.}
  \ee %
The conformal extensions of  \BdS/\BAdS-space can be realized via
     the conformal extensions of
  \dS/\AdS-hyperboloid $H_{\pm}$, respectively, first and back to
  the Beltrami coordinates
  afterwards.

  Introducing a scaling variable $\kappa\neq0$ and
    a set of coordinates $\zeta^{\hat A}$, $\hat A=0,\cdots, 5$,
  respectively,
  \be\label{Liexi+}%
    \zeta^i := \kappa \xi^i, ~
    \zeta^4 := \kappa \xi^4, ~
    \zeta^5:=\kappa R, \quad {\rm for} ~H_{R +};\\
      \zeta^i := \kappa \xi^i, ~
    \zeta^4 := \kappa R, ~
    \zeta^5 := \kappa \xi^4, \quad {\rm for} ~H_{R -}.
    \label{Liexi-}
  \ee
  Then, under such a scaling, eq.~(\ref{5sphr}) turns out to be %
  \begin{equation}
   \mathcal{N}:~ \eta_{\hat A \hat B} \zeta^{\hat A} \zeta^{\hat B}
    = 0, \quad \eta_{\hat A \hat B} = diag(J^{1,3}, -1,1),
  \label{LieS0}
  \end{equation}
  where $J^{1,3}=diag(1,-1,-1,-1)$, ${\hat \zeta} := (\zeta^{\hat{A}}) = (\zeta, \zeta^4,
  \zeta^5) \neq 0$.
  This is an $SO(2,4)$-invariant null cone $\mathcal{N} \subset
 M^{2,4}$ and there is the projective equivalence relation $\sim$ on $M^{2,4} - \{0\}$:
  ${\hat \zeta}' \sim {\hat \zeta}$ if and only if there is a number $c\neq 0$ satisfying
  $\zeta'^{\hat{A}} = c \ \zeta^{\hat A}$.
  The resulted quotient space $[\cal N] := {\cal N}/\!\!\sim$ is a 4-dimensional
  submanifold of $\mathbb{R}P^5$, homeomorphic to $S^1 \times S^3$.
  Intuitively,  an equivalence class of ${\hat \zeta} \in {\cal N}$ can be viewed as
  the null straight line passing through both ${\hat \zeta}$ and the origin of $M^{2,4}$.
  The origin is not included in the equivalence class, however.
  In this sense, $[\mathcal{N}]$ consists of all the null straight lines
  through the origin.  Thus, an $SO(2,4)/\mathbb{Z}_2$ transformation on $M^{2,4}$ induces
  a transformation on $[{\cal N}]$.

Since $H_{\pm}$ can be embedded into $[\mathcal{N}]$,
  when the metric on $M^{2,4}$ is pulled back to $[\mathcal{N}]$, it is
  conformal to $ds^2_{H_\theta}$ in (\ref{ds2pm}):
   \be\label{ds2C}%
    d\chi^2|_{[\mathcal{N}]} : =
    \eta_{\hat{A}\hat{B}}\,d\zeta^{\hat{A}}\,d\zeta^{\hat{B}}|_{[\mathcal{N}]}=
    \kappa^2 \ ds^2_{H_\theta}.
  \ee %
 Consequentially, an $SO(2,4)/\mathbb{Z}_2$  transformation on $[{\cal N}]$  induces the conformal
  transformation on $H_{\pm}$, respectively:
  \begin{eqnarray}\label{cdspm}%
     ds'^2_{H_\pm}
    = \rho^2\,ds^2_{H_\pm},\quad
     \rho = \frac{\kappa}{\kappa'}
    = \left\{
      \begin{array}{ll}
        {\zeta^5}/{\zeta'^5}, \quad & \textrm{for } H_{R+} \\
        {\zeta^4}/{\zeta'^4}, \quad & \textrm{for } H_{R-}.
      \end{array}
      \right .
  \end{eqnarray}

According to eqs.~(\ref{Liexi+}) and (\ref{Liexi-}), $H_\pm$ can be
viewed as an intersection of $\mathcal{N}$ and the hyperplanes $P_+:
\zeta^5 = R$ and $P_-: \zeta^4 = R$, respectively.  Since $H_\pm$ is
only part of $\mathcal{N}$ with $\zeta^5\neq 0$ for $H_+$ or
$\zeta^4\neq 0$ for $H_-$, respectively, it is quite possible for an
$SO(2,4)$ transformation to send a point in $H_\pm$, with nonzero
$\zeta^5$ or $\zeta^4$, to another one with zero $\zeta^5$ or
$\zeta^4$, and vice versa.  Thus, $H_\pm$ are, in fact, not closed
under the induced conformal transformations.  To be closed, $H_\pm$
must be extended into the whole $[{\cal N}]$.  Thus, $[{\cal N}]$ is
the conformal extension of both \dS/\AdS-spaces.

 It is clear that back to the Beltrami atlas, say (\ref{u4})
for the \BdS, as inhomogeneous projective
  coordinates, the conformal \BdS/\BAdS-metric follows. It is straightforward to
  prove that all null geodesics
of the   conformal \BdS/\BAdS-metric are straight world lines,
respectively. Thus, we get  conformal extensions of \dS/\AdS\,
\SR\ for those massless particles and light signals on
\BdS/\BAdS-spacetime, respectively. Both them are defined on the
same $[\cal N]$.

 As is well-known,  the conformal \Mink-space can also be obtained
  from the the same null cone (see, e.g.,  \cite{twistor}). To this end, it is needed
   to introduce a set of new coordinates
  \begin{equation}
    \zeta^{\pm} :=\frac{1}{\sqrt{2}} (\zeta^5\pm\zeta^4)
  \end{equation}
  and inhomogeneous projective coordinates 
  \be\label{Mcdt}%
    x^i := R\,\zeta^i/\zeta^-, \quad
    x^+ := R\,\zeta^+/\zeta^-
  \ee
  to those points with $\zeta^- \neq 0$, where $R$ is the same universal constant
  introduced before. In general, different $R$ may  be taken.
  Then eq.~(\ref{LieS0}) becomes
  $x^+ = - \eta_{ij}\,x^i x^j/(2R)$,
  and  metric (\ref{ds2C}) becomes
  \be %
    d\chi^2|_{[\cal N]} = (\zeta^-/R)^2 \ ds_M^2,&& %
    ds_M^2 := \eta_{ij} \ dx^i dx^j. %
    \label{ihN}
  \ee%
  Now an $SO(2,4)/\mathbb{Z}_2$ transformation on $[{\cal N}]$ induces a
  conformal transformation on
  \Mink-space:
  \be\label{cmink}%
    ds^2_M \rightarrow ds'^2_M = \rho^2\,ds^2_M, \quad \rho=
    \zeta^-/\zeta'^-.
  \ee%

  Similarly, the \Mink-space  can be regarded as an intersection of
  ${\cal N}$ and the hyperplane $P_M: \zeta^- = R$ by identifying $(x^i)$
  with $(x^i, (x^+ - R)/\sqrt{2}, (x^+ + R)/\sqrt{2}) \in {\cal N}$.  The
  \Mink-space is also not closed for these conformal transformations.
  Thus, the \Mink-space needs to be extended, resulting in the space
  $[{\cal N}] \cong S^1\times S^3$.


  According to the  above discussion, the \Mink/\dS/\AdS-space and their
  conformal extensions  can be
  related by Weyl conformal maps.  An event on $dS$, say, is first viewed
  as an event on $P_+\cap\mathcal{N}$.  Then  an event  on $P_-\cap\mathcal{N}$
  equivalent to it could be found, in general.
  However, it is possible that  an event on one space could not be mapped into
  another space, or could not find an inverse image on another space.
  But, this can be solved so that  the map from the conformal extension of $dS$
  to that of $AdS$ can be established. We will explain this issue in
  detail elsewhere.

  For example,  an event  $\xi_+ := (\xi_+^0, \ldots, \xi_+^4) \in H_+$
  can be mapped to  an event  on $H_-$ with  following Beltrami coordinates:
  \begin{equation}
    x_-^i := R\,\zeta^i/\zeta^5  = \xi_+^i.
    \label{dS2AdS}
  \end{equation}
  As another example, the {Weyl conformal map} sending  an event  with
  coordinates $(x^i)$ on the
  \Mink-space to  an event  on the \BdS-space with  coordinates $(x^i_+)$ reads
  \begin{equation}
    x^i_+ = - \sqrt{2}\,x^i\,
      ({1 + \frac{1}{2R^2}\,\eta_{jk}\,x^j x^k})^{-1}.
    \label{M2dS}
  \end{equation}
  This is just the conformally flat coordinate transformation for
  the \BdS-metric (\ref{metric}) also known as a stereographic
  projection with an inverse transformation
  \begin{equation}
    x^i = - \sqrt{2}\,x_+^i ({1 \mp \sqrt{\sigma(x_+)}})^{-1}.
    \label{dS2M}
  \end{equation}
  The sign $\mp$ is opposite to the sign of $\xi^4_+\gtrless 0$ in the \BdS-space.

  It is important that the normal vectors of $P_+$, $P_-$ and
  $P_M$ are time-like, space-like and null, respectively, and that
  $P_+\cap\mathcal{N}$, $P_-\cap\mathcal{N}$ and $P_M\cap\mathcal{N}$
  is
  \dS/\AdS/\Mink-space, respectively.  This can be generalized: given a
  hyperplane off the origin, its intersection with ${\cal N}$ is \dS/\AdS/\Mink-space
   if its normal vector is time-like, space-like or null, respectively.

%
%
\subsection{Triality of null physics on conformal \Mink/\dS/\AdS-spaces}

 We have shown that \Mink/\dS/\AdS-spaces can all be conformally extended to
  the same $[\mathcal{N}]$, so that they can be conformally mapped from one
  to another via Weyl mappings.
  A conformal transformation on one space is, in fact, also a conformal
  transformation on another space.  And all these conformal transformations are
  induced from some transformations of
  $SO(2,4)/\mathbb{Z}_2$, due to the equivalence relation on $\mathcal{N}$.
  Therefore, from the viewpoint  of conformal transformations, these three
  kinds of spaces and  \CFT s on them are just same.  We refer to this
  fact as a triality of conformal extensions of these spaces and
  the null physics on them including 
  the
   \AdS/\CFT-correspondence.
   Thus, 
there should be also a triality for the conjecture. %

\subsubsection{Motion of free massless particles and light signals}
\label{sect:nullGeod}

 As was mentioned, similar to a massive particle a free massless
  particle or a light signal in  \dS-spacetime is in the uniform `great circular' motion
  with   a  conserved
  5-d angular momentum (\ref{L5}). In terms of the Beltrami coordinates,
  the uniform `great circular' motion turns out to be inertial motion along a null
  straight line \cite{BdS, BdS2}. These imply that a geodesic is the intersection of
  $\Sigma$ and \dS-hyperboloid $H_+$ in (\ref{5sphr}), where $\Sigma$ is some 2-d plane passing through the origin
  of the 5-d \Mink-space $M^{1,4}$. 
   It can be proved that, when the geodesic is null, it
  is in fact a straight line in $M^{1,4}$, having the equation $\xi^A = \xi^A_0
  + \lambda_+\,v^A$ for some constants $\xi^A_0$ and $v^A$, satisfying
  $\eta_{AB} \xi^A_0 v^B = \eta_{AB} v^A v^B = 0$.  Thus, the 5-d momentum
  $K^A = v^A$ of the null geodesic is also conserved.

  Using the relations~(\ref{Liexi+}),
  we can obtain
  \begin{equation}
    \label{Ls}
    L^{AB} = \frac{1}{\kappa^2}\,\frac{d\psi}{d\lambda}\,\mathcal{L}^{AB},
    \quad
    P^A = \frac{1}{\kappa^2 R^2}\,\frac{d\psi}{d\lambda}\,\mathcal{L}^{5A},
  \end{equation}
  where $\psi = \psi(\lambda)$ is a certain  parameter and
the 6-d angular momentum $\mathcal{L}^{\hat{A}\hat{B}}$ is defined
as
  \begin{equation}
    {\cal L}^{\hat{A}\hat{B}}
    := \zeta^{\hat{A}}\,\frac{d\zeta^{\hat{B}}}{d\psi}
    - \zeta^{\hat{B}}\,\frac{d\zeta^{\hat{A}}}{d\psi}.
    \label{6-angmomentum}
  \end{equation}
It is conserved if
  \begin{equation}
    d\psi=\kappa^2{d\lambda}.
    \label{psi-lambda}
  \end{equation}
 For a massless particle in the \AdS-space, there are similar issues.

  In the \Mink-case, the 4-d momentum $k_M^i$ and the  angular momentum
  $l_M^{ij}$ are conserved for a light signal
  \begin{equation}
    k_M^i := \frac{dx^i}{d\lambda}, \quad
    l_M^{ij} := x^i k_M^j - x^j k_M^i.
  \end{equation}
  Similarly, there is  a 6-d angular momentum
  \begin{eqnarray}
    && \mathcal{L}^{ij} =
    \frac{d\lambda}{d\psi}\,\kappa^2\,l_M^{ij}\,\quad
    \mathcal{L}^{4j}
    = \frac{1}{\sqrt{2}}\,(\mathcal{L}^{+j} - \mathcal{L}^{-j}),
    \\
    && \mathcal{L}^{5j}
    = \frac{1}{\sqrt{2}}\,(\mathcal{L}^{+j} + \mathcal{L}^{-j}),
    \quad
    \mathcal{L}^{45} = \mathcal{L}^{+-},
  \end{eqnarray}
  where %
  \be\label{6-L}%
    \mathcal{L}^{-j} = \frac{d\lambda}{d\psi}\,\kappa^2
    R\,P^j,\quad
   \mathcal{L}^{+j} = \frac{d\lambda}{d\psi}\,\kappa^2\,
      (x^+ P^j - x^j \frac{dx^+}{d\lambda} ), \quad
      \mathcal{L}^{+-} = - \frac{d\lambda}{d\psi}\,\kappa^2\,R\,
      \frac{dx^+}{d\lambda}.
     \ee%
  If eq.~(\ref{psi-lambda}) is satisfied, then the above 6-d angular momentum
  is also conserved.

  For a massless free particle, its equation of motion in $[{\cal N}]$ is not
  unique in terms of $\zeta^{\hat{A}}$, because
  $\zeta^{\hat{A}} = \zeta^{\hat{A}}(\psi)$ and
  $
    \zeta^{\hat{A}} = \zeta'^{\hat{A}}(\psi')
    := \rho(\psi')\,\zeta^{\hat{A}} (\psi(\psi'))
  $
  are equivalent, with $\psi = \psi(\psi')$ a re-parameterization.
  Formally, there are the angular momenta $\mathcal{L}^{\hat{A}\hat{B}}$ and
  $\mathcal{L}'^{\hat{A}\hat{B}}(\psi')$ for the same particle.
  But, a re-parameterization can always be chosen so that
  $\mathcal{L}'^{\hat{A}\hat{B}}(\psi')$ is still conserved.

  Consequently, the world-line is lying in a 2-d plane $\Sigma$ passing through
  the origin of $M^{2,4}$, which is also contained in \NsM\ except for the
  origin.  Thus, the world-line $\Sigma - \{0\}/\!\!\sim$ is a projective
  straight line in $[\mathcal{N}]$: in the Beltrami coordinate  on \dS/\AdS, or in
  \Mink-coordinate, its equations look like
  \begin{equation}
    x^i(s) = x^i_0 + \tau\,c^i,
  \end{equation}
  where $x^i_0$ and $c^i$ are some constants while $\tau$ is the curve
  parameter.  Hence, the world-line is a null geodesic \cite{BdS, BdS2}.
  The relation of its 5-d angular momentum and ${\cal L}^{\hat{A}\hat{B}}$ is
  as shown in  eqs.~(\ref{Ls}), etc.
  This coincides with the well known fact that null geodesics are conformally
  invariant up to a re-parameterization.

%
%

\subsubsection{On \CFT\ and  \AdS/\CFT\ correspondence}
\label{sect:CFT}

Let us consider other conformal issues and their relations on
conformal \Mink/\dS/\AdS-spaces.

The generators of the conformal group on  \Mink-space are
\begin{eqnarray}
  && \hat{p}_i := \partial_i, \qquad
  \hat{l}_{ij} := x_i \ \partial_j - x_j\,\partial_i,
  \\
  && \hat{D} := x^l \ \partial_l,  \quad
  \hat{s}_i := - x\cdot x \,\partial_i + 2 x_i x^l\,\partial_l.
\end{eqnarray}
A \CFT\ on \Mink-space must be invariant under  action of these
generators.  Coordinates $x^i$ can be extended to be a set of
coordinates $(x^i, \kappa, \phi)$ on $M^{2,4} - \{\zeta^- = 0\}$,
where $\kappa$ is the scaling factor introduced before
\begin{eqnarray}
  \kappa = \frac{\zeta^-}{R}, \quad
  \phi := \eta_{\hat{A}\hat{B}}\,\zeta^{\hat{A}} \zeta^{\hat{B}}.
\end{eqnarray}
Thus, the \Mink-space is described by $\kappa = 1$ and $\phi = 0$.
Then it can be verified that
\begin{eqnarray}
  && \hat{p}_i = \frac{1}{R}\,\hat{\mathcal{L}}_{+i}, \quad
  \hat{l}_{ij} = \hat{\mathcal{L}}_{ij},
  \label{Poincare} \\
  && \hat{D} = \hat{\mathcal{D}} + \hat{\mathcal{L}}_{-+}, \quad
  \hat{s}_i = 2x_i\,\hat{\mathcal{D}}
  + 2R\,\hat{\mathcal{L}}_{-i},
  \label{SK}
\end{eqnarray}
where
\begin{equation}
  \hat{\mathcal{D}} := \zeta^{\hat{A}}\,
    \frac{\partial}{\partial\zeta^{\hat{A}}}
\end{equation}
is the generator of scaling in $M^{2,4}$, while
\begin{equation}\label{gso24}
  \hat {\cal L}_{\hat A \hat B}
  := \zeta_{\hat A} \ \frac{\partial}{\partial \zeta^{\hat B}}
  - \zeta_{\hat B} \ \frac{\partial}{\partial\zeta^{\hat A}}
\end{equation}
are generators of  \cg\ (up to a fact $i$).  Since
$\hat{\mathcal{D}}$ is commutative with each
$\hat{\mathcal{L}}_{\hat{A}\hat{B}}$, it does not matter that the
conformal generators of the \Mink-space differ from those of
$M^{2,4}$ by a vector field along $\hat{\mathcal{D}}$ (see,
eqs.~(\ref{SK})).  This coincides with (i) the idea that the
equivalence relation $\sim$ will be considered on $\mathcal{N}$, and
(ii) the fact that  conformal transformations on the \Mink-space are
induced from, but not the same as, \CG-transformations on
$\mathcal{N}$.  In fact, a quantity on the \Mink-space can be
realized by homogeneous function of degree zero on $M^{2,4} -
\{0\}$. In this way $\hat{\mathcal{D}}$ somehow could be dropped
directly.

Generators of conformal transformations on \dS/\AdS-spaces, or
specially on \BdS/\BAdS-spaces, can also be given as the ones of
\cg.  Thus, they can be related by the Weyl conformal mappings such
as (\ref{dS2AdS}) and (\ref{M2dS}). Correspondingly, the \CFT s in
these spaces are also related by these mappings. Since the Maxwell
equations are the simplest \CFT, as an illustration, we show how the
sourceless Maxwell equations
\begin{equation}
  d \vect{F} = 0, \qquad *\  d *\vect{F} = 0,
  \label{Maxwell}
\end{equation}
where $*$ is the Hodge dual operator, are related among them.

Consider  the  Weyl conformal mapping $\psi: M^{1,3} \rightarrow
dS^4$ as shown in eq.~(\ref{M2dS}):
\begin{equation}
  \psi^* \vect{g} = \Omega^2\,\vect{\eta}, \qquad
  \Omega = \sqrt{2} \ \Big( 1 - \frac{1}{2R^2}\,\eta_{ij}\,x^i x^j
    \Big)^{-1},
\end{equation}
with  $\vect{g}$ the metric (\ref{metric}) of $BdS^4$, $\vect{\eta}$
the one in (\ref{ihN}).  If $\vect{F}_{dS}$ is the Maxwell field on
$dS$, its equations follow
\begin{equation}
  d \vect{F}_{dS} = 0, \qquad
  \star\ d \star\vect{F}_{dS} = 0,
  \label{Maxwell-dS}
\end{equation}
where $\star$ is the dual operator with respect to $\vect{g}$.  We
pull $\vect{F}_{dS}$ back to the \Mink-space, resulting in
\begin{equation}
  \vect{F} = \psi^* \vect{F}_{dS}.
  \label{eq:F}
\end{equation}
Thus, $d\vect{F} = d \ (\psi^*\vect{F}_{dS}) = \psi^* d\vect{F}_{dS}
= 0$ is satisfied. It can be verified that
\begin{eqnarray*}
  \psi^*(\star\,d\star\vect{F}_{dS})
  = \Omega^{-2}\,[*\,d * \vect{F}].
\end{eqnarray*}
Therefore, on the \Mink-space, $\vect{F}$ as in eq.~(\ref{eq:F}) is
a sourceless electromagnetic field: eqs.~(\ref{Maxwell}) are
satisfied.  In this way the Weyl conformal mapping $\psi: M^{1,3}
\rightarrow BdS^4$ relates a sourceless electromagnetic field
$\vect{F}_{dS}$ on the \BdS-space to a sourceless $\vect{F}$ on the
\Mink-space.

Similarly, this approach can be applied to other \CFT s between \dS\
and \AdS-spaces, \AdS\ and \Mink-spaces and so on.  Basically, the
\CFT s on \Mink/\dS/\AdS-spaces, in which all the relevant fields
are assumed to behave well as the infinity points are approached,
can be unified together.  The former is merely a realization of the
latter.

For the \AdS/\CFT\ correspondence, there should also be a triality.

A 5-dimensional $AdS$-space with radius $R_5$ can be embedded into
$M^{2,4}$ as a
hypersurface $\mathcal{S}$:%
 \be%
  {\cal S}:~ \eta_{\hat{A}\hat{B}}\,\zeta^{\hat{A}} \zeta^{\hat{B}} =
  R_5^2.
\ee%
 If antipodal points on $\mathcal{S}$ are identified, the
resulted space, denoted  $\mathcal{S}/\mathbb{Z}_2$, is still
homoemorphic to $\mathcal{S}\cong AdS^5$.  In the projective space
$\mathbb{R}P^5 = M^{2,4} - \{0\}/\sim$, the quotient space of those
$\zeta^{\hat{A}}$ satisfying $\eta_{\hat{A}\hat{B}}\,
\zeta^{\hat{A}} \zeta^{\hat{B}} > 0$ are  homeomorphic to
$\mathcal{S}/\mathbb{Z}_2 \cong AdS^5$.  Identifying $AdS^5$ with
this quotient space,  its boundary is just the null cone modulo a
projective equivalence
\begin{equation}\label{boundary-ads5}
  \partial_P(AdS^5) \cong [\mathcal{N}].
\end{equation}

Thus, due to the triality of the \CFT s in conformal
\Mink/\dS/\AdS-spaces, there should be three {\AdS/\CFT}
correspondences starting from the well-known \AdS/\CFT\
correspondence \cite{adscft}. Namely, there should be  the
{\AdS/\CFT} correspondence between $AdS^5$ and  $dS^4$/$AdS^4$,
respectively, in addition to that between $AdS^5$ and  \Mink-space.
Clearly, this triality of the \AdS/\CFT\ correspondence can be
generalized to any dimensions whenever the \AdS/\CFT\ correspondence
is conjectured.

%
%

\section{Theory of Gravity with Localization of Maximum Symmetry}\label{gravity}


 In this section, we  explain why gravity should be based on the localization of
  \SR\ with full maximum symmetry and be governed by some  gauge-like dynamics
 of the same local maximum symmetry. We  also construct a kind of
 umbilical manifolds with local \dS-invariance and briefly
 introduce a simple model of  \dS-gravity with a gauge-like
 dynamics characterized by a dimensionless coupling constant $g \cong 
({G\hbar
 \Lambda}/ {{3}c^{3}})^{1/2}\sim 10^{-61}$. Although this model is
 quite simple, it may still shed light on why our universe is so
 dark.

\subsection{From the equivalence principle to  the principle of localization}

  As was quoted before, right after
 explain why  there is  `an argument in a circle' for the \PoI\ and raised a severe question
 on the existence of inertial systems,  Einstein claimed that `\omits{
 the principle of
 inertia as  established, to a high degree of approximation, $\cdots$, provided that we neglect the perturbations due to the sun and
 planets. Stated more exactly, }there are finite regions, $\cdots$
 in which the laws of the special theory of
 relativity $\cdots$ hold with remarkable accuracy. Such regions
 we  shall call ``Galilean regions".' \cite{1923} Then Einstein explained why the spacetimes
 with gravity should be curved. This is the most remarkable and most
 successful point of view in Einstein's \GR, although his argument on rotating
  disc is
 fallacious.

 Let us analyze Einstein's above statement from both physical
 and geometrical  viewpoints.

 Firstly, since  all these regions are `{\it
  finite}', `{\it in which the laws of the special theory of
 relativity, $\cdots$, hold with remarkable accuracy},' it is important to note
 that the Poincar\'e symmetry of
 the laws of  special
 relativity 
 on these `finite
  regions'   should be eventually {\it local}.  Although in practice, Poincar\'e symmetry in these regions
   may still be regarded as {\it global symmetry} approximately.

Secondly, let us consider  how to pass from one  `Galilean
region'
to another  at different but   nearby
 positions in the spacetime with gravity and  what
 kind of local symmetry should be for the curved spacetime with gravity.
 According to Einstein, there should be gravity
 in-between
 these `regions'. Therefore, in order to transit from one to another, some
 paths on curved
 spacetime with gravity in-between should be passed. \omits{In other words, in order to
 connect  these
  `regions' together,
 some gravitational field as
 interaction  should be taken into account. }Since there is local Poincar\'e symmetry
 in  these `regions', in order to transit along these paths in-between,
 the curved spacetime with gravity should
 also be of some local symmetry.
 It would be better still the local Poincar\'e symmetry.
 Otherwise, it is
 hard to transit consistently  from one `region' to another
 if Poincar\'e  symmetry cannot be maintained  locally
 in the course of transition along certain path in-between.
 For any number of such `finite
 regions', it is the same.

This may also be seen from another angle more mathematically. Each
of the {\it finite} `Galilean regions' is essentially a portion of
 a \Mink-space with Poincar\'e symmetry isomorphic to an $R^4$, so
 that
 there are intersections among these
  \Mink-spaces with different  `finite regions' at different positions and
  the transition functions on these
 intersections  should also be
 valued in Poincar\'e symmetry. Further, in terminology  of differential geometry,
 these \Mink-spaces with `finite regions' may be
 viewed as tangent spaces  at different positions of a curved
  manifold as the spacetime with
 gravity and the transition functions in the intersections of different coordinate charts
 on the manifold should be valued in {\it local} Poincar\'e symmetry.

 Thus, it is the core of  Einstein's idea on gravity  
 that the theory of gravity should be based on the
localization of his \SR\ with full Poincar\'e  symmetry anywhere
and anytime on some curved spacetimes. For the sake of
definiteness, we name this principle
as the  local \PoR\ 
or {\it the principle of localization}.

 Since there are three kinds of \SR\  of Poincar\'e/\dS/\AdS-invariance, 
   there should
 be also three kinds of gravitational effects with full
local Poincar\'e/\dS/\AdS-symmetry, respectively. The {\it \PoL\,}
 states: {\it On spacetimes with gravity, there always exist
local relativity-frames of local \Mink/\dS/\AdS-spacetime,
physical laws must take the gauge covariant versions of their
special-relativistic forms with respect to the local
Poincar\'e/\dS/\AdS-symmetry, respectively.}

In \GR, however, the \PoE\ 
requires: `In any and every local Lorentz frame, anywhere and
anytime in the universe, all the (nongravitational) laws of physics
must take on their familiar special-relativistic forms.' \cite{MTW}.
It is clear that on ${3+1}$-dimensional pseudo-Riemannian geometry
$({M}, {\mathbf g})$ with metric $\bf g$ of signature $-2$ as
spacetime with gravity, there is no local translation symmetry in
local Lorentz space as tangent space (see, e.g., \cite{KN} and for
some earliest references, see, e.g., \cite{Cartan}). Actually, the
definitions for mass, spin and other physical quantities of
particles and fields as test objects or gravitational sources as
well as the physical laws they obeyed in \GR\ are merely made
formally `on their familiar special-relativistic forms' in local
Lorentz frames. As far as the local symmetry is concerned in \GR, it
is $GL(4,R)$ or its subgroup $SO(1,3)$.

For example, a
rank-$(r,s)$ tensor $T(x)$ is defined as at a point %
\be\label{rstensor}%
T(x):=T^{i_1,\cdots,
i_r}_{~~~~~~~j_1,\cdots,j_s}(x)\frac{\partial}{\partial
x^{i_1}}\otimes \cdots\otimes \frac{\partial}{\partial
x^{i_r}}\otimes
dx^{j_1}\otimes\cdots\otimes dx^{j_s}.%
\ee%
In the same coordinate chart, it is invariant under the
transformations of bases of the tangent space and its dual, i.e.
$(\frac{\partial x^{'i}}{\partial x^j})_{i,j=0,\cdots,3} \in GL(4,
R)$ at the point. It is also invariant from one chart to another
on an intersection of two charts since  transition functions are
also valued in $GL(4,R)$.

In Einstein's special relativity, however, the full Poincar\'e
symmetry plays a central role for the \PoI\ as the benchmark for
physics. In fact, the mass and the spin, which characterize systems
invariant under Poincar\'e group \cite{wigner}, are related to the
eigenvalues of two Casimir operators, in which
 translation generators always appear, of
Poincar\'e algebra $\mathfrak{iso}(1,3)$:%
\be\label{CsmP}%
C_1:=\eta^{jk}{\hat p}_j{\hat p}_k,  \qquad
C_2:=\eta^{jk}{\hat w}_j{\hat w}_k,%
\ee%
where ${\hat w}_j:=\epsilon_{jklm}{\hat p}^k {\hat l}^{lm}$ is the
Pauli-Lubanski vector, ${\hat p}^j:=\eta^{jk}{\hat p}_k, {\hat
l}^{lm}:=\eta^{lr}\eta^{ms}{\hat l}_{rs}$, ${\hat p}_j, {\hat
l}_{jk}$ generators of translations and homogeneous Lorentz algebra
$\mathfrak{so}(1,3)$, respectively. It was Wigner \cite{wigner} who
found that although 
 spin also corresponds to the rotation group symmetry $SU(2)$ as
a subgroup of homogeneous Lorentz group $SO(1,3)$, {\it but only
if $m^2>0$}. In the case $m^2=0$, the spin is no longer described
by $SU(2)$ and this, in fact, is why the polarization states of  a
massless particle with spin $s$ are $s_z=\pm s$ only. For example,
physical photons do not exist in a $s_z=0$ state, whereas massive
spin $1$ particles do (see, e.g.,  \cite{ryder}). This is also the
case that there is no longitudinal component   for the
electromagnetic wave in the vacuum.

Thus, the benchmarks for physics in Einstein's \SR\ and \GR\ seem to
be not completely in consistency with each other in symmetry and its
localization. This
 may lead to some potential problems.
In order to get rid of this kind of problems, it is reasonable to
require an enhanced equivalence principle with localization of
\SR\ of full symmetry, {\it the \PoL,} as was proposed above.
\omits{ in the light of Einstein's `Galilean regions'.In other
words, there should be {\it the localized principle of relativity}
or {\it the
 principle of localization}: {\it On spacetimes with gravity, there
always exist local relativity-frames with local Poincar\'e, \dS\ or
\AdS-symmetry, physical laws   must take the gauge covariant
versions of their special-relativistic forms with respect to the
local Poincar\'e, \dS\ or \AdS-symmetry, respectively.}}

How to describe the general spacetimes  with gravity based upon
the \PoL?

As was mentioned earlier, firstly, $\cal M$ should be a kind of
3+1-dimensional manifolds with
 metric $\bf g$ of local relativity-frames  in  corresponding special relativity.
 Secondly,  in order to describe
 that there is localized full symmetry in the corresponding
special relativity at each event on $\cal M$,  a kind of bundles
$E( {\cal M}, {\cal S}, {\mathfrak G}, P)$ is needed with  $\cal
M$ as base manifold, the maximally symmetric spacetime $\cal S$,
one of the \Mink/\dS/\AdS-spacetimes, as typical fibre and the
maximum symmetry $\mathfrak G$, one of $ISO(1,3)/SO(1,4)/SO(2,3)$,
as structure group. And there  should be also a principal bundle
$P({\cal M}, {\mathfrak G})$. Thirdly, gravity with localized full
symmetry should be described by the matric $\bf g$ or its local
frames and a kind of connections $\Gamma$ valued in the Lie
algebra $\mathfrak g$ of  $\mathfrak G$. It is important that in
principle these bundles with required connections can be
constructed.

As was mentioned, however, the  pseudo-Riemann manifolds
with local Lorentz frames in \GR\ is just a  special case: the
bundle $E({M}, M^{1,3}, G, P)$ with   pseudo-Riemann manifold $M$
as base manifold and the \Mink-spacetime $M^{1,3}$ as  fibre. It
is clear that such a geometrical description is not complete from
the viewpoint of  the \PoL, since the structure group $G$ is just $GL(4,R)$
or its subgroup $SO(1,3)$.

\subsection{Principle of localization and gravitational dynamics}

In \GR, Einstein-Hilbert equation  reads symbolically \cite{MTW}%
\be\label{Eeq}%
{\bf G}=8\pi G{\bf T},%
\ee%
where ${\bf G}$ is  Einstein tensor, ${\bf T}$  energy-momentum
tensor of sourse and $G$  Newton's gravitational constant. The
Einstein-Cartan `moment of rotation' $\bf G$ \cite{MTW} is made of
Riemann-Christoffel curvature. From the  viewpoint of  holonomy theorem,
however, the curvature is basically related to local homogeneous
Lorentz rotation (see, e.g., \cite{trautm, gtg, guo79}). But, $\bf
T$ is in a same form with the stress-energy tensor related to the
translation invariance of matter on the \Mink-spacetime in  view
of  Noether's theorem (see, e.g., \cite{trautm,  gtg, guo79}).

Although by means of variational principle,  Einstein-Cartan `moment
of rotation' $\bf G$  is derived from
variation of Einstein-Hilbert action 
with respect to  metric or  coefficients of Lorentz frame, which
may be regarded as a kind of `translation' connection from the  viewpoint
 of
Cartan's structure equation or as canonical affine connection (see,
e.g., \cite{KN}). Thus, it seems more or less still reasonable to
connect it with the stress-energy tensor $\bf T$, which is also
given by the variation of the matter's action with respect to the
same variable(s),  metric or  coefficients of Lorentz frame.
However, in  connection theory (see, e.g., \cite{KN}),
the coefficients of Lorentz frame can be regarded as a kind of
`translation' connection for what is called  the canonical affine
connection. Namely, there should be an affine structure locally on
the spacetimes with gravity. This is just in consistency with the
\PoL\ with respect to Poincar\'e invariance. Therefore, the
spacetimes with gravity should be in general pseudo-Riemann-Cartan
manifolds with torsion rather than pseudo-Riemann manifolds
without torsion.

On the other hand, a spinning particle with mass $m$ moves 
with a curvature-spinning current force in \GR\ \cite{
MTW, weinberg}: %
\be\label{sforce}%
m\frac{D^2x^k}{ds^2}=f R_{ab}^{~~kl}S^{ab}_{~~l},%
\ee%
where $R_{ab}^{~~kl}:=e_a^i e_b^j R_{ij}^{~~kl}$, $e_a^i$
coefficients of Lorentz frame, $R_{ij}^{~~kl}$ Riemann curvature,
 $S^{ab}_{~~l}$ spinning
current of the particle and $f$ a free parameter. It is important to
note that although $f$ may be very tiny, the coupling is like the
Lorentz-force of a charged particle moving in electromagnetic
field, which is of gauge coupling. Therefore, in \GR\ there are
two kinds of couplings between  gravity and matter: The one in
Einstein-Hilbert equation (\ref{Eeq}) and that in (\ref{sforce}).

Thus,  some questions can be raised: Why does the dynamics connect
geometry with matter in different (local) symmetry in field
equation? Why  gravitational fields should not be described by the
both curvature and torsion? Why the spinning current as a property
of the matter with respect to  spacetime symmetry does undergo an
action from  curvature as gravity, but cannot effect  gravity as a
kind of source?

Cartan suggested that  Einstein-Hilbert equation should be
generalized by what is called Einstein-Cartan equations now
\cite{Cartan,  trautm, gtg, guo79, EC, held}, which
read symbolically :
\be\label{ECeq}%
{{\bf G}_\Gamma}=8\pi G{\bf T},\qquad {\bf Y}=8\pi G {\bf S},%
\ee%
where $ {\bf G}_\Gamma$ is  Einstein-like tensor of  Cartan's
connection $\Gamma$ or $B^{ab}_{~j}\in \mathfrak{so}(1,3)$, ${\bf
Y}$  con-torsion of the connection and ${\bf S}$  spin-current of
gravitational source\omits{, respectively (for  precise
expressions of these objects, it is referred to  appendix)}.
However, there is still another kind of gauge-like coupling in the
equation of motion for test (spinning) particles. Thus, from the viewpoint
of  holonomy theorem and  Noether's theorem, the questions on
connect between geometric quantities and physical quantities are
still there \cite{ gtg, guo79}.

According to the \PoL, it seems reasonable to require further that 
geometry and matter should be connected in same local symmetry and the
gravitational dynamics  be of local invariance of the \PoL.
Namely, the gravitational dynamics should  be in consistency with
the \PoL. This also indicates that gravitational field equations
be of gauge-like with localized symmetry of the \PoL\ (see, e.g.
\cite{gtg, guo79}). Of course, correct equations should pass
observation tests for \GR\ at least.

\omits{We shall show that these points have been indicated  by a
simple model of \dS-gravity on a kind of umbilical manifolds of
local \dS-invariance in a special gauge.}

\subsection{Localization of \dS-hyperboloid and umbilical manifold}

Simply speaking, the spacetimes with gravity of local
\dS-invariance may be described as a kind of $3+1$-dimensional
umbilical manifolds ${\cal M}^{1,3}:={\cal H}^{1,3}$ as
sub-manifolds of $4+1$-dimensional manifolds ${\cal M}^{1,4}$.
This reflects a localization of the \dS-hyperboloid \HsM\,
\cite{uml}.

Let us illustrate how to construct such an ${\cal M}^{1,3}:={\cal
H}^{1,3} \subset {\cal M}^{1,4}$.

 Suppose
there is an local \HsM\ anywhere and anytime tangent to the ${\cal
M}^{1,4}$ such that at a point $p \in {\cal H}^{1,3}$, the radius
vector ${\bf r}_p$ with norm $R$ of the \HsM\ is oppositely normal
to the tangent \Mink-space of ${\cal H}^{1,3}$, i.e. ${\bf r}_p=-{
N}_p$, at the point. Since this local \Mink-space is also tangent to
the \HsM\ at the point, which  is umbilical for the \HsM\ in ${\cal
M}^{1,4}$. Thus, ${\cal H}^{1,3}$ consists of all these points,
which are umbilical in the above sense, and is a sub-manifold of the
${\cal M}^{1,4}$, i.e., ${\cal H}^{1,3} \subset {\cal M}^{1,4}$.
Such a kind of Riemann-Cantan manifolds ${\cal H}^{1,3}$ are called
umbilical manifolds with an umbilical structure of \HsM\ anywhere
and anytime.

This construction can also be given in an opposite manner: Given a
point $p$ on ${\cal H}^{1,3}$, there is a local \Mink-space as the
tangent space at the point, $T_p({\cal H}^{1,3})$, and given a
vector ${(N=Rn)}_p$ of norm $R$ at the point with an $n_p$ as the
unit base of space $N^1_p$ normal to $T_p({\cal H}^{1,3})$ with a
metric of \dS-signature in ${\cal M}^{1,4}$. Then the space $T_p
\times {N }^1_p \cong M^{1,4}_p$ is tangent to ${\cal M}^{1,4}$ at
the point. Thus, under local \dS-transformations on $T_p \times {N
}^1_p \cong M^{1,4}_p$ there is a local hyperboloid structure $H_R
\subset M^{1,4}_p$ isomorphic to the \dS-hyperboloid \HsM\ in
(\ref{H+}) at the point $p$ as long as $R{n}_p=-{\bf r}_p$ is taken.
In fact, all these points consist of the umbilical manifold ${\cal
M}^{1,3}:={\cal H}^{1,3} \subset {\cal M}^{1,4}$.

Therefore, on the co-tangent space $T_p^*$ at the point $p \in {\cal
H}^{1,3}$ there is a Lorentz frame 1-form:
\be\label{Lframe}%
\theta^b=e^b_jdx^j, ~~\theta^b(\partial_j)=e^b_j; \quad
e^a_je^j_b=\delta^a_b, ~~~e^a_je_a^k=\delta^k_j; %
\ee%
with respect to a Lorentz inner product:%
\be\label{Lpro}%
<\partial_j,\partial_k>=g_{jk}, ~~<e_a, e_b>=\eta_{ab},
~~\eta_{ab}=diag(1,-1,-1,-1). %
\ee%
Here, $\partial_j$ the base of the  tangent space $T_p$. 
 The line-element on ${\cal H}^{1,3}$
can be expressed as%
\be%
ds^2=g_{jk}dx^jdx^k=\eta_{ab}\theta^a\theta^b,
\quad~g_{jk}=\eta_{ab}e^a_je^b_k.%
 \ee%

 There is a Lorentz covariant derivative a la Cartan:
 \be%
 \nabla_{e_a}e_b=\theta^c_{~b}(e_a)e_c; \quad %
 \theta^a_{~b}=B^a_{~b j}dx^j,\quad~\theta^a_{~b}(\partial_j)=B^a_{~bj}. %
\ee%
$B^a_{~cj}\in \mathfrak{so}(1,3)$ are connection coefficients of
\omits{. In order to keep the inner product being invariant, }the
Lorentz connection 1-form
$\theta^{ab}=\eta^{bc}\theta^a_{~c}$. 
The torsion and curvature can be defined as%
\be\nno%
\Omega^a&=&d\theta^a+\theta^a_{~b}
\wedge\theta^b=\frac{1}{2}T^a_{~jk}dx^j\wedge dx^k\\\label{T2form}
&&T^a_{~jk}=\partial_je^a_k-\partial_ke^a_j+B^a_{~c j}e^c_k-B^a_{~c
k}e^c_j;\\\nno%
\Omega^a_{~b}&=&d\theta^a_{~b}+\theta^a_{~c}\wedge\theta^c_{~b}=\frac{1}{2}F^a_{~b
jk}dx^j\wedge dx^k\\\label{F2form}
&&F^a_{~b jk}=\partial_jB^a_{~bk} -\partial_kB^a_{~bj}+B^a_{~cj}B^c_{~bk}-B^a_{~ck}B^c_{~bj}.%
\ee%
 They satisfy  corresponding Bianchi identities.%
\omits{\be\label{BianchiL1}
d\Omega^a&=&\Omega^a_{~c}\wedge\theta^c-\theta^a_{~c}\wedge
\Omega^c,\\\label{BianchiL2} %
d\Omega^a_{~b}&=&d\Omega^a_{~c}\wedge
\theta^c_{~b}-\theta^a_{~c}\wedge\Omega^c_{~b}. \ee}

It is easy to get a metric compatible affine connection
$\Gamma^i_{~jk}$ from the
requirement%
\be\label{De=0}%
g_{jk/l}=0, ~\Leftrightarrow~e^a_{~j//k}=0=\partial_k
e^a_{~j}-\Gamma^i_{~jk}e^a_{~i}+B^a_{~ck}e^c_{~j}.%
\ee%

 As was just mentioned, at the point $p \in {\cal
H}^{1,3}$, there are a space $N^1_p$ and its dual ${N^1_p}^*$
normal to ${\cal H}^{1,3}$ with a normal vector $n$ and its dual $
\nu$ on $T_p({\cal M}^{1,4})$ and $T^*_p({\cal M}^{1,4})$,
respectively. Namely, $\{\partial_j, n; dx^i,\nu\}$ and $\{e_a,n;
\theta^b, \nu\}$ span $M_p^{1,4}=T_p^{1,3}\times N_p^1$ and
${M_p^{1,4}}^*={T_p^{1,3}}^*\times {N_p^1}^*$, respectively. Let
these
bases satisfy the following conditions in addition to (\ref{Lpro})%
\be\label{dSpro}%
dx^i(n)=\theta^b(n)=0,&&\nu(\partial_j)=\nu(e_a)=0,~~~n(\nu)=1;\\
<e_a,n>=0,&& <n,n>=-1.%
\ee %
Then, the \dS-Lorentz base $\{\hat{E}_A\}$ and their dual
$\{\hat\Theta^B\}$ can be defined as: %
\be\label{dSL}%
\{\hat{E}_A\}=\{e_a, n\},~~\{\hat{\Theta}^B\}=\{\theta^b, \nu\}.%
\ee%
And (\ref{Lpro}) and (\ref{dSpro}) can be expressed as%
\be\label{dSLpro}%
\hat\Theta^B(\hat E_A)=\delta^B_A, ~~<\hat E_A,
\hat E_B>=(\eta_{A B})_{A,B=0,\cdots, 4}=diag(1,-1,-1,-1,-1).%
\ee%
Introduce a normal vector $N=Rn$ with norm $R$:%
\be\label{N}%
N=Rn=\hat \xi^A \hat E_A, ~~(\hat \xi^A)=(0,0,0,0,R), ~~<N,N>=-R^2.%
\ee%
For  the \dS-Lorentz base, there are 
\be\label{LocalS}%
g_{jk}=\eta_{AB}\hat E^A_j \hat E^B_k,~~\eta_{AB}\hat\xi^A\hat
E^b_j=0,~~\eta_{AB}\hat \xi^A\hat\xi^B=-R^2,%
 \ee%
where %
\be%
\hat E^A_j=\hat \Theta^A(\partial_j), ~~\{\hat
E^A_j\}=\{{ e^a}_j, 0\}. %
\ee%

The transformations, which maps $M^{1,4}_p$ to itself and
preserves the inner product, are
\be%
\hat E_A\rightarrow E_A=S^B_A \hat E_B, ~~\hat \Theta^A\rightarrow
\Theta={S^{-1}}^A_B \hat \Theta^B, ~~S{J}S^t={J},%
\ee%
where ${J}=(\eta_{AB})=diag(1,-1,-1,-1, -1),~S=(S^A_B)\in
SO(1,4)$, $*^t$ denotes the transpose. The transformed base is
defined as the \dS-base and its dual
$E_A, \Theta^B$, respectively: 
\be\label{dSB}%
\Theta^A(E_B)=\delta^A_B,&&\Theta^A(\partial_j)=E^A_j,\quad~
<E_A,E_B>=\eta_{AB}.\\\label{LocalS1}
g_{jk}=\eta_{AB}E^A_jE^B_k,&&\eta_{AB}\xi^A
E^B_j=0,\quad~\eta_{AB}\xi^A\xi^B=-R^2,%
\ee%
where $E^B_j$ are the \dS-frame coefficients. Obviously, these
formulas reflect the local \dS-invariance on ${\cal H}^{1,3}$ and
(\ref{LocalS1}) show that there is a local 4-dimensional
hyperboloid $H^{1,3}_p\subset M^{1,4}_p$ tangent to ${\cal
H}^{1,3}$ at the point $p$. Thus, (\ref{LocalS1}) may be called
{\it the local \dS-hyperboloid condition}.


Now  the \dS-covariant derivative a la Cartan can be introduced%
\be\label{dSCD}%
\hat \nabla_{E_A}E_B=\Theta^C_{~B}(E_A)E_C.%
\ee%
$\Theta^A_{~C}\in \mathfrak{so}(1,4)$ is the \dS-connection
1-form. In the local
coordinate chart $\{x^j\}$,%
\be\label{dSconnection}%
\hat \nabla_{\partial_j}E_B=\Theta^C_{~B}(\partial_j)E_C={B}^C_{~Bj}E_C,%
\ee %
${B}^A_{~Cj}$ denote the \dS-connection coefficients. There are
also the \dS-torsion ${\bf \Omega}^A$, curvature 2-forms ${\bf
\Omega}^A_{~B}$ and  their Bianchi identities.


In the light of  Gauss formula and Weingarten formula in the surface
theory \cite{Spivak}, from the \dS-covariant derivative of the
\dS-Lorentz base (\ref{dSL}) with  properties of $\theta^a,
~\theta^a_b$, it follows a
generalization of  Gauss formula and Weingarten formula
\be\label{dSLGauge}%
\hat\nabla_{\partial_j}e_a=\theta^b_{~a}(\partial_j)e_b-b_{ab}\theta^b(\partial_j)n,
~~\hat\nabla_{\partial_j}n=b^a_{~b}\theta^b(\partial_j)e_a.%
\ee%
 Here, $b_{ab}$ denotes a
second fundamental form of the hypersurface.
Since ${\cal H}^{1,3}$ is supposed to be an umbilical 
hypersurface, where every point satisfies the umbilical condition on
\HrsM%
\be\label{umbilic}%
g_{jk}=Rb_{jk},%
\ee%
these formulas read on ${\cal H}^{1,3}$%
\be\label{dSLGaugeS}%
\hat\nabla_{\partial_j}e_a=\theta^b_{~a}(\partial_j)e_b-R^{-1}\theta_a(\partial_j)n,
~~\hat\nabla_{\partial_j}n=R^{-1}\theta^a(\partial_j)e_a.%
\ee%
On the other hand, for the \dS-Lorentz base from
(\ref{dSCD}) there are%
\be\label{dSLCD}%
\hat\nabla_{\partial_j}e_a=\check
{\Theta}^b_{~a}(\partial_j)e_b+\check{\Theta}^4_{~a}(\partial_j)n,
~~\hat\nabla_{\partial_j}n=\check{\Theta}^a_{~4}(\partial_j)e_a,%
\ee%
where $\check{\Theta}$ denotes the \dS-connection ${\Theta}$ in
the \dS-Lorentz gauge.

Comparing with (\ref{dSLGaugeS}), it follows%
\be\label{dSLconnect}%
\check{\Theta}^{ab}(\partial_j)={\theta}^{ab}(\partial_j)=B^{ab}_{~~j},
&&~\check{\Theta}^{a4}(\partial_j)=R^{-1}{\theta}^{a4}(\partial_j)=R^{-1}e^a_j; \\\nno%
\check{\cal B}^{ab}_{~~j}=B^{ab}_{~~j}, &&~\check{\cal B}^{a4}_{~~j}=R^{-1}e^a_j.
\ee%
Namely, the \dS-connection in the \dS-Lorentz gauge may be written as%
\be\label{dSc}%
(\check {B}^{AB}_{~~j})=\left(
\begin{array}{cc}
B^{ab}_{~~j} & R^{-1} e^a_{~j}\\
-R^{-1}e^b_{~j} &0
\end{array}
\right ) \in \mathfrak{so}(1,4).
\ee%
This is just the connection  introduced in \cite{dSG, uml, T77,
QG}. Here, it is recovered from the umbilical manifolds with local
\dS-invariance.

\omits{
Further,  on ${\cal M}^{1,3}:={\cal H}^{1,3}$, there is a
connection valued in \dS-algebra in a special gauge made of the
Lorentz
 connection and the Lorentz frame:
\be\label{dSLB}%
{\check{\cal B}}:={\check{\cal B}}_jdx^j, ~ {\check{\cal B}}_j:=(
{\check{\cal B}}^{AB}_{~~j})_{A,B=0,\cdots, 4}= \left(
\begin{array}{cc}
B^{ab}_{~~j} & R^{-1} e^a_j\\
-R^{-1}e^b_{j} &0
\end{array}
\right ) \in \mathfrak{so}(1,4), %
\ee%
where $\check{\cal B}$ denotes the connection is in a special
gauge called the Lorentz gauge since the rest gauge freedom is of
Lorentz algebra. }The corresponding curvature reads:
\be\label{dSLF}%
{\check{\cal F}}:=d{\check{\cal B}}+{\check{\cal B}}\wedge
{\check{\cal B}}=\frac{1}{2}{\check{\cal F}}_{jk} dx^j \wedge
dx^k,\qquad\qquad\qquad\quad\\\nno%
 {\check{\cal F}}_{jk}= ( {\check{\cal
F}}^{AB}_{~~jk})=\left(
\begin{array}{cc}
F^{ab}_{~~jk} + 2R^{-2}e^{ab}_{~~ jk} & R^{-1} T^a_{~jk}\\
-R^{-1}T^b_{~jk} &0
\end{array}
\right ) \in \mathfrak{so}(1,4),
\ee%
where $e^a_{~bjk}=\frac{1}{2}(e^a_je_{bk}-e^a_ke_{bj}),
e_{bj}=\eta_{ab}e^a_j$, $ F^{ab}_{~~ jk}$ and $ T^a_{~jk}$ are
curvature (\ref{F2form}) and torsion (\ref{T2form}). 

\subsection{A simple model of \dS-gravity}%
 For the \dS-connection (\ref{dSc}), a simple model of
 \dS-gravity can be introduced \cite{dSG, uml,
T77, QG}.

The total action of the model 
with source may be taken as%
\be\label{S_t}%
S_T=S_{GYM}+S_m,%
\ee%
where $S_m$ is   action of  source with minimum coupling, and
$S_{GYM}$ the Yang-Mills-like action of the model as follows (in
the \dS-Lorentz gauge):
\be\nno%
S_{GYM}&=&\frac{\hbar}{4g^2}\int_{{\cal M}^{1,3}}d^4x { e}
{\bf Tr}_{dS}({\check {\cal F}}_{jk}{\check {\cal F}}^{jk})\\
&=& \int_{{\cal M}^{1,3}}d^4x { e}
 \left [  \d {c^3} {16\pi G} (F-2\La) -\d {
\hbar}{4g^2} F^{ab}_{~~\mu\nu} F_{ab}^{~~\mu\nu}+ \frac {c^3} {32\pi
G} T^{a}_{~\mu\nu}T_a^{~\mu\nu} \right ].\label{GYM}
\ee%
Here $e=\det(e^a_j)$, a dimensionless constant $g$ should be
introduced as usual in the gauge theory to describe the
self-interaction of the gauge field,\omits{ $\chi$ a dimensional
coupling constant related to $g$ and $R$, and} $F={
\frac{1}{2}}F^{ab}_{~jk}e_{ab}^{~jk}$ the scalar curvature of Cartan
connection, the same as the action in  Einstein-Cartan theory. In
order to make sense in comparing with  Einstein-Cartan theory, we
should take 
$g^2 \cong 
{G\hbar
 \Lambda}/ {{3}c^{3}}\sim 10^{-122}$.\omits{, the same as the one introduced in
 the last section
in the sense of the Planck scale-$\Lambda$ duality. This is why we
have used the same symbol in the different cases.}

\omits{The field equations can be given  via the variational
principle with respect to $e^a_{~j},B^{ab}_{~~j}$:\omits{the
gravitational equations with
sources that are different from the schematic eq. (\ref{Geq})}%
\be\label{Geq2}
{T^a_{~jk ||}}^k &+& F^a_{~j}-\frac{1}{2}F e^a_{~j}{\blue
convention ~of ~both}-{\blue ?}\Lambda
e^a_{~j}={\red +}-8\pi G( T^a_{m j}{\red -}+T^a_{G j}),{\red sign ??\blue ?}\\\label{Geq2'}
{F^{ab}_{~jk||}}^k &=& -8\pi G(S^{ab}_{mj}+S^{ab}_{G j}),%
\ee
$||$:  The double covariant derivative w.r.t.  $\{\}, \omega{\blue
?}$; arbitrary coordinate transformations and  Lorentz rotation;
$F_{jk}=F^a_{~j}e_{a k}$: Ricci-like tensor w.r.t. the Lorentz
curvature, $T^a_{m j}, T^a_{G k}$: e-m tensor for matter and gravity;
$S^{ab}_{mj}, S^{ab}_{Gj }$: spin current for  matter and gravity.

Especially, the e-m-like tensor $T^a_{G j}$ for gravity
\be\label{emG}
T^a_{G j}&:=&g^{-2} T^a_{F j}+2\chi T^a_{T j},\quad
T^a_{*~j}=T_{*kj}e^{a
k}\\\label{emF}
T_{F jk}&:=&-\frac{1}{\varepsilon} \frac{\delta(\varepsilon{Tr
F^2})} {\delta g^{jk}}
=Tr(F_{~jl}F_{k}^{~l})-\frac{1}{4}g_{jk}Tr(F_{lm}
F^{lm}),\\\label{emT}
T_{T jk}&:=&
T^a_{~jl}T_{a k}^{~l}-\frac{1}{4}g_{jk}T^a_{~lm}T_a^{~lm},%
\ee%
$T_{F jk}$: e-m-like tensor for  curvature; 
                      $T_{T jk}$: e-m-like tensor for  torsion.
 The spin-like currents for  gravitational field are given by%
\be\label{spG}
S^{ab}_{G j}&=&S^{ab}_{F j}+S^{ab}_{Tj},\\\nonumber%
S^{ab}_{F j}&:=& 
Y^l_{~jk}e^{ab k}_{~~~l}-Y^{l k}_{~~k}
e^{ab}_{~~jl}{\red 2nd ~term?},\\\nno%
S^{ab}_{T j}&:=& T^{[a}_{~jk}e^{b]k},
\ee%

$S^{ab}_{Fj}$: 
the spin-like current for  curvature $F$; 
$S^{ab}_{Tj}$: the spin-like current for 
torsion $T$. }

 It {is} natural to see that the gravitational field
equations now should be of gauge-like. But, different from
ordinary gauge theory, there is some energy-momentum-like
tensor $T^a_{G j}$ for gravity itself as source from  variation with
respect to the coefficients of Lorentz frame of the third and last
term in the action (\ref{GYM}), respectively:
\be%
T^a_{G j}&:=&{ g^{-2}}T^a_{F j}+{ 2\chi}T^a_{T j},\quad
T^a_{*~j}=T_{*kj}e^{a k}, ~~\chi= c^{3}/{G\hbar
 }\label{emG}\\
T_{F jk}&:=&{\rm Tr}(F_{~jl}F_{k}^{~l})-\frac{1}{4}g_{jk}{\rm
Tr}(F_{lm} F^{lm}),\label{emF}\\
T_{T jk}&:=& T^a_{~jl}T_{a
k}^{~l}-\frac{1}{4}g_{jk}T^a_{~lm}T_a^{~lm},\label{emT}
\ee%

For  the case of spinless for  matter and torsion-free for
gravity, the field equations
become  Einstein-Yang equations \cite{gwz} with $\Lambda$-term (in what follows, we
take unit of $c=\hbar=1$).%
\be%
R^a_{~j}-\frac{1}{2}e^a_{~j}R+\Lambda e^a_{~j}&=&-8\pi G(T^a_{m
j}+{ g^{-2}}T^a_{R j}),\label{Geq3}\\
{R^{ab}_{~~jk ||}}^k&=&0,\label{Geq3'}
\ee%
where $||$ is the double covariant derivative with respect to
Christoffel and Ricci rotation coefficients $\gamma^{ab}_{~j}$,
$T^a_{m j}=e^{a k}T_{m jk}$ the energy-momentum tensor of matter,
and $ T^a_{R j}=e^{a k}T_{R jk}$  the energy-momentum-like tensor of
Riemann curvature $R^{ab}_{~~jk}\in \mathfrak{so}(1,3)$
\be\nno%
T_{R j}^{~~k}
&=&{R}_{ab jl}{R}^{abkl} - \frac{1}{4}\dl_j^k({R}_{ablm}{R}^{ablm}),\nno \\
&=&2C_{lj}^{~~mk}{R}^l_m +\frac{{R}}{3}({R}_j^k
-\frac{1}{4}{R}\dl_j^k), \label{emR}
\ee%
where $C_{ljmk}$ is  Weyl tensor. For the last equation in
(\ref{emR}), the G\'eheniau-Debever decomposition for  Riemann
curvature is used. It is clear that if  Ricci tensor vanishes, i.e.,
$R_{jk}=0$, this energy-momentum-like tensor of Riemann curvature
(\ref{emR}) vanishes so that the vacuum solutions in \GR\ do satisfy
the Einstein-Yang equations (\ref{Geq3}) and (\ref{Geq3'}) without
$\Lambda$-term \cite{wzc}.

It is easy to prove that for  \dS-spacetime the
`energy-momentum'-like tensor in (\ref{emR}) vanishes { as well,} so
\dS-spacetime also satisfies  eqs (\ref{Geq3}) and (\ref{Geq3'}).
\omits{Therefore, the 4-dimensional Riemann sphere is just a
gravitational instanton in this model. }It can also be proved that
all solutions of vacuum Einstein equation with $\Lambda$-term do
satisfy these equations, so this simple model does pass the
observation tests in solar-scale. Further, it is shown \cite{hg}
that some simplest cosmic models may have `Big Bang' but differ from
\GR, as $T_{R jk}$ could play a role as a kind of the `dark stuffs'.
Since the general equations are of gauge-like, there are
gravitational potential waves of the both metric and Cartan's
connection including the gravitational metric waves in \GR.

It is important that  the \dS-gravity in this  model is
characterized by a dimensionless coupling constant $g$ like in
ordinary gauge theory. This is one of reasons why the model is
renormalizable \cite{QG}. It is also interesting that it is of an
$SO(5)$ gauge-like Euclidean action with the Riemann sphere being
an instanton. Thus, the quantum tunneling scenario may support
$\Lambda>0$. For the gauge-like gravity, asymptotic freedom may
indicate the coupling constant $g$ \omits{should be running to
zero at infinity momentum. For gravity, however,  , so $g$ }should
be very tiny and link the cosmological constant $\Lambda$ with the
Planck length $\ell_P$ properly, since $\Lambda$ and $\ell_P$ as
 a fixed point should provide an infrared and an ultraviolet cut-off,
 respectively \cite{duality, dual07}.

This model presents some important indications to why the universe
is so dark. First, the cosmological constant $\Lambda$  as a
fundamental constant is introduced from the `gauge' symmetry so that
it is not just a `dummy' constant at classical level put in by hand
in \GR. And it should play a role of the simplest dark energy. In addition,
there are some candidates for the dark matter from \dS-gravity
itself, such as the `energy-momentum-like tensors' for gravity and
so on. In fact, by means of the relation between Cartan's connection
$B^{ab}_{~j}$ and Ricci rotational coefficients $\gamma^{ab}_{~j}$,
Einstein-Hilbert action can be picked up from the first term in
(\ref{GYM}), all other terms except the cosmological constant
$\Lambda$, which is the simplest form of the dark energy,  are all
the dark matter from the  viewpoint of \GR. Thus, this model should provide an
alternative framework for the dark-data analysis in precise
cosmology.

\section{Concluding  Remarks}

In the last century physics, symmetry, its localization and
symmetry breaking play very important roles. For  physics  in the
large scale, it should be also the case. Namely, the maximally
symmetric spacetime with maximum symmetry and their localization
should play a central role.

Initiated by Professor Lu's proposal \cite{Lu}, there are  three
kinds of special relativity \cite{Lu, LZG, lu05, BdS, BdS2, BdS3,
IWR, TdS, NH, yan, duality, OoI, PoI} based on the \PoR\ on
\dS/\AdS-spacetimes, or Poincar\'e \PoR\ as its \Mink-contraction
$R\to \infty$, and the postulate on invariant universal constants,
or its \Mink-contraction. All other kinematics with the \PoR\,
should be their contractions \cite{NH}. \omits{Further, relevant
gravity should be based on the principle of localization with
gauge-like dynamics, respectively \cite{duality, OoI, PoI}.}

From the viewpoint of the \dS\ \SR, the dark energy is at least mainly
the cosmological constant $\Lambda$ and 
 \dS-spacetime provides an important model:  There is the
\PoR\ and a
 law of inertia 
in Beltrami coordinate atlas  with  Beltrami simultaneity. The
proper-time simultaneity flips it to another side of 
a Robertson-Walker-like \dS-space with an accelerated expanding
$S^3$ fitting the cosmological principle.\omits{Thus, the
Robertson-Walker-like cosmos  acts as the origin of the inertial
law.} If our universe is asymptotic to such a Robertson-Walker-like
\dS-space, it should be slightly closed in $O(\Lambda)$ with
$R\backsimeq (3/\Lambda)^{1/2}$ and all celestial objects including
the CMB in the cosmic scale should be rotated qualitatively. On the
other hand, the universe can fix on Beltrami systems  via its
evolution. Therefore, for the \PoI\ on \dS-spacetime and its all
contractions there should be no Einstein's `argument in a circle'
\cite{PoI} and the universe just acts as the origin of inertia
\cite{BdS3, OoI, PoI}.

For  null physics of three kinds of special relativity,  symmetry
should be enlarged to  conformal group realized on the same
 projective  null cone  isomorphic to the projective boundary
 of a 5-dimensional \AdS-space, i.e.,  $[{\cal N}]\cong\partial_P(AdS^5)\subset
M^{2,4}$. \omits{extensions of the \PoI\ can be made. In fact, the
conformal extension of \dS/\AdS\ \SR\ can be made in addition to
that of Einstein's \SR\ on the same null space $[{\cal N}]\subset
M^{2,4}$ with $SO(2,4)/\mathbb{Z}_2$ invariance, which is the
boundary of a 5-dimensional \AdS-spacetime. }Thus, there is a
triality for conformal extensions of  null physics on
\Mink/\dS/\AdS-spacetimes including the  \AdS/\CFT\ correspondence.
And there should be a \dS-spacetime on the boundary of $S^5 \times
AdS^5$ as a vacuum of supergravity.

Gravity should be based on the principle of localization  with
localized \PoR\ of full maximum symmetry. Thus, the localization of
\SR\ leads to corresponding theory of gravity with local maximum
symmetry. For \dS-gravity, its dynamics should be
 gauge-like in consistency  with the \PoL\ characterized by a dimensionless constant $g\simeq
(\Lambda G\hbar
/3c^{3})^{1/2}\sim 10^{-61}$. 
 A simple model \cite{dSG, uml, T77, QG}
 shows the
features on a kind of umbilical Riemann-Cartan manifolds of local
 \dS-invariance \cite{uml}. Some  gravitational effects in this model that cannot be
 included
 in \GR\ should play the role as the dark matter. 

What are the benchmarks for physics? Whether these benchmarks are
consistent  each other? These are most important and fundamental
issues.

If the \PoR\,
should be generalized to all maximally symmetric spacetimes \omits{
such as  \dS-spacetime of radius $R\simeq(3/\Lambda)^{1/2}$ and its
\AdS-counterpart or the \Mink-spacetime of $R\to \infty$; and as in
Newton's mechanics and Einstein's \SR, the \PoI\ with maximum
symmetry and be regarded as the benchmark of physics without
gravity. Further, }and if  gravity should be described based on the localized \PoI\
 with full maximum
symmetries, \omits{ so as to the localized \PoI\  be the benchmark of physics
on spacetimes with gravity, respectively.
  Thus, }the benchmark for
physics with gravity is in consistency with the one without gravity
of  \SR\ \cite{PoI}. 

Some seventy years ago,  Einstein claimed: `Physics constitutes a
logical system of thought which is in a state of evolution'.
`Evolution is proceeding in the direction of increasing simplicity
of the logical basis (principles).' `We must always be ready to
change these notions - that is to say, the axiomatic basis of
physics - in order to do justice to perceived facts in the most
perfect way logically.' \cite{1936} This has greatly enlightened
us how to understand  evolution of physics in the past and how to
look forward the direction of its evolution. Especially, how to
deal with the theory of relativity as a kind of `principle theory'
in the face of the challenges from the dark universe.

It seems that study on special relativity and theory of gravity
via maximum symmetry and its localization is  in a right
direction: increasing simplicity of the principles, `in order to
do justice to perceived facts in the most perfect way logically'.

The dark universe and its asymptotic behavior may already indicate
that the \dS\ \SR\ and the \dS-gravity 
 should be the foundation  of physics in the large scale.

\omits{Symmetry and its localization play extreme important roles
in physics.

There should be three kinds of relativity and their contractions
\cite{NH}, and three kinds of theory of gravity as localization of
corresponding relativity with a gauge-like dynamics.

Via {interchangeable} relation between Snyder's-like quantized
space-time models in \dS/\AdS-space of momenta and the
\dS/\AdS-invariant special relativity on \dS/\AdS-spacetime with the
Beltrami-Kliein coordinates, respectively, there should also be  a
duality in the physics at and between the Planck scale and
cosmological constant bridged by  gravity based on the localization
of corresponding \SR\ with a gauge-like dynamics of full localized
symmetry characterized by a dimensionless coupling constant.

{The dark universe indicates the \dS\ \SR\  and  gravity with local
\dS-invariance should be the foundation of physics in large scale.}}

\begin{acknowledgments}
We would like to thank Professor Q.K. Lu as well as
 Z. Chang, C.-G. Huang, W.L. Huang, J.Z. Pan,  X.C. Song, Y. Tian, R.S. Tung, S.K. Wang,
  F. Wu, K.
Wu, X.N. Wu, Z. Xu, M.L. Yan, X. Zhang, B. Zhou, C.J. Zhu and Z.L.
Zou for valuable discussions or collaborations. This work is partly
supported by NSFC under Grant No. 90503002.
\end{acknowledgments}

\end{document}